\documentclass[oupdraft]{bio}
\usepackage{url}

\usepackage{amssymb}
\usepackage{amsmath}
\usepackage{graphicx}
\usepackage{algorithm}
\usepackage{algpseudocode}
\usepackage{bbm}
\usepackage{natbib}

\newtheorem{cctheorem}{Theorem}

\newtheorem{newlemma}{Lemma}
\newtheorem{condition}{Condition}

\history{Received August 1, 2010;
	revised October 1, 2010;
	accepted for publication November 1, 2010}

\begin{document}
	
	% Title of paper
	\title{\vspace{-15mm}Unlocking the Power of Time-Since-Infection Models: Data Augmentation for Improved Instantaneous Reproduction Number Estimation}
	
	\author{JIASHENG SHI, YIZHAO ZHOU, 	JING HUANG$^\ast$\\
		\textit{Department of Biostatistics, Epidemiology, and Informatics, Perelman School of Medicine, University of Pennsylvania, Philadelphia, USA}
		\\
		% E-mail address for correspondence
		{jing14@pennmedicine.upenn.edu}}
	
	% Running headers of paper:
	\markboth%
	% First field is the short list of authors
	{J. Shi and others}
	% Second field is the short title of the paper
	{Data Augmentation for Improved Instantaneous Reproduction Number Estimation}
	
	\maketitle
	
	% Add a footnote for the corresponding author if one has been
	% identified in the author list
	\footnotetext{To whom correspondence should be addressed.}

\begin{abstract}
{The Time Since Infection (TSI) models, which use disease surveillance data to model infectious diseases, have become increasingly popular due to their flexibility and capacity to address complex disease control questions. However, a notable limitation of TSI models is their primary reliance on incidence data. Even when hospitalization data are available, existing TSI models have not been crafted to improve the estimation of disease transmission or to estimate hospitalization-related parameters - metrics crucial for understanding a pandemic and planning hospital resources. Moreover, their dependence on reported infection data makes them vulnerable to variations in data quality. In this study, we advance TSI models by integrating hospitalization data, marking a significant step forward in modeling with TSI models. We introduce hospitalization propensity parameters to jointly model incidence and hospitalization data. We use a composite likelihood function to accommodate complex data structure and an Monte Carlo expectation–maximization algorithm to estimate model parameters. We analyze COVID-19 data to estimate disease transmission, assess risk factor impacts, and calculate hospitalization propensity. Our model improves the accuracy of estimating the instantaneous reproduction number in TSI models, particularly when hospitalization data is of higher quality than incidence data. It enables the estimation of key infectious disease parameters without relying on contact tracing data and provides a foundation for integrating TSI models with other infectious disease models.}
{Composite likelihood function; Hospitalization propensity; Infectious disease transmission; Time between diagnosis and hospitalization.}
\end{abstract}
\vspace{-6mm}
\section{Introduction}
\label{sec:intro}
Mathematical modeling is essential for understanding infectious disease transmission, especially during the early stages of pandemics when vaccines, treatments, and immunity are limited. Policymakers rely on key metrics like the instantaneous reproduction number ($R_t$), projected incidence, and hospitalization counts to make timely decisions. Various modeling approaches, including compartmental models, agent-based simulations, and time-series analyses, help estimate these metrics. Among these, Time-Since-Infection (TSI) models have gained prominence due to their ability to estimate $R_t$ directly from incidence data using renewal equation concepts.

The concept of TSI models originated in the 1910s by \cite{ross1916application} and \cite{ross1917application_a, ross1917application_b} and was mathematically formalized by  \cite{kermack1927contribution}. They gained further prominence with contributions from \cite{wallinga2004} and \cite{fraser2007estimating}. Recognition in the statistics community grew after the work of \cite{cori2013new} and the development of the EpiEstim R package \citep{EpiEstim}, which enabled widespread use for real-time $R_t$ estimation, particularly during the COVID-19 pandemic \citep{pan2020association,gostic2020practical,nash2022real}. Recent enhancements include regression integrations, addressing delays, and multi-location analyses \citep{quick2021,shi2022robust,ge2023effects}. These models are based on the idea that new infections depend on three factors: recent infection counts, the current reproduction number, and how transmission evolves over time. They are mathematically linked to renewal equations, which model new cases as a sum of past infections weighted by infectious potential, and have similarities with branching processes that describe how each case leads to secondary infections \citep{athreya2004branching,lloyd2005superspreading}. Tools like the EpiNow2 package \citep{EpiNow2} have applied these concepts to estimate time-varying transmission dynamics, accommodating various infectiousness patterns and reporting delays.

%Here, we provide a brief overview of the fundamentals of TSI models. In practice, consider a discrete-time setting where $I_t$ represents the number of new infections on day $t$. Given the infection history up to day $t-1$, the expected new infections on day $t$ can be expressed as $E(I_t | I_0, \ldots, I_{t-1}) = R_t \Lambda_t$, where $\Lambda_t = \sum_{s=1}^t I_{t-s} \omega_s$ represents the infection potential on day $t$. This potential is shaped by the current number of infectious individuals and the infectiousness function $\omega_s$, which describes how infectious an individual is $s$ days after infection, with $\sum_{s=0}^{+\infty} \omega_s = 1$. Typically, $\omega_s$ is set to zero for $s = 0$ and for $s > \eta$, where $\eta$ is the time between infection and recovery. This function is often approximated using the distribution of the serial interval or generation time \citep{svensson2007note}. Estimating $R_t$ involves assuming a distribution for $I_t$ and applying methods like maximum likelihood or Bayesian approaches with predefined values of $\omega_s$ \citep{cori2013new,who2014ebola,quick2021}.

Compared to other infectious disease models, e.g., compartmental models, TSI models are valued for their practicality and flexibility. Their statistical foundation enables integration with advanced methods, making them well-suited for complex modeling challenges. By relying primarily on empirical data, TSI models require minimal epidemiological knowledge, making them particularly useful during new infectious disease outbreaks with limited biological insights \citep{jewell2021statistical}. A key advantage is their ability to estimate $R_t$ directly from incidence data alone, making them user-friendly during early pandemic stages when other surveillance data are scarce. However, this reliance on incidence data can introduce biases from underreporting or reporting delays. Moreover, when other crucial data, such as hospitalization counts, become available, traditional TSI models are not equipped to integrate this information. It is because they assume each new infection is linked to prior cases, a pattern not followed by hospitalization data. This limitation restricts TSI models from effectively modeling disease-related hospitalizations, an essential metric for assessing pandemic trajectory and planning healthcare resources.

This study aims to address these limitations by developing an enhanced TSI model that integrates hospitalization data. We investigate whether incorporating hospitalization data can improve the precision of $R_t$ estimates and offer insights into disease dynamics and severity in TSI models. Specifically, our approach aims to: 1) reduce bias in $R_t$ estimation from incidence data inaccuracies, 2) estimate hospitalization-related parameters for assessing disease severity, and 3) combine the strengths of TSI and compartmental models for pandemic response. To achieve this, we introduce a new set of parameters, the hospitalization propensity, which quantifies the tendency of infectious individuals being hospitalized over time since infection. By incorporating hospitalization data into the TSI framework, we use a composite likelihood function to jointly model incidence and hospitalization counts, capturing their interdependent dynamics \citep{lindsay1988composite,varin2011overview}. The composite likelihood approach simplifies the modeling of complex data dependencies, allowing us to retain key statistical properties while reducing computational challenges. Using a Monte Carlo expectation–maximization (MCEM) algorithm, which combines stochastic sampling with iterative parameter estimation to handle missing data, we estimate model parameters and assess their accuracy through simulations and an analysis of COVID-19 data across U.S. counties. Our results demonstrate the enhanced precision of $R_t$ estimation and highlight the capability of the proposed framework in supporting hospital resource planning and evaluating local transmission risks.

This work contributes to the field by unlocking the potential of TSI models through data augmentation of incidence and hospitalization data, which refines real-time estimation of $R_t$, enables the estimation of hospitalization-related parameters previously accessible primarily through contact tracing, and facilitates analysis of associations between risk factors and disease transmission dynamics. The sections that follow outline the methodology, implementation strategy, and broader implications for infectious disease modeling.

\vspace{-5mm}

\section{Methods}
\label{s:model}

We first provide a brief overview of the fundamentals of TSI models, and then describe the proposed approach to integrate incidence data with hospitalization data based on the TSI framework.

\vspace{-6mm}
\subsection{Model and Notations}
\label{model}

We use subscripts $t$ to denote calendar time and $s$ to indicate time since infection. Consider a discrete-time setting where $I_t$ represents the number of new infections on day $t$. Given the infection history up to day $t-1$, the expected new infections on day $t$ can be expressed as $E(I_t | I_0, \ldots, I_{t-1}) = R_t \Lambda_t$, where $\Lambda_t = \sum_{s=1}^t I_{t-s} \omega_s$ represents the infection potential on day $t$. This potential is shaped by the current number of infectious individuals and the infectiousness function $\omega_s$, which describes how infectious an individual is $s$ days after infection, with $\sum_{s=0}^{+\infty} \omega_s = 1$. Typically, $\omega_s$ is set to zero for $s = 0$ and for $s > \eta$, where $\eta$ is the time between infection and recovery. This function is often approximated using the distribution of the serial interval or generation time \citep{svensson2007note}, and estimating $R_t$ involves assuming a distribution for $I_t$ and applying methods like maximum likelihood or Bayesian approaches with predefined values of $\omega_s$ \citep{cori2013new,who2014ebola,quick2021}.

%We use subscripts $t$ to denote calendar time and $s$ to indicate time since infection. 
 By ``hospitalization", we refer to hospital admission due to the infectious disease. Let $H_t$ denote the number of new hospitalizations on day $t$. We define the filtrations $\mathcal{F}_t = \sigma ( \{ I_r, 0\leq r\leq t \} )$ and $\mathcal{G}_t =  \sigma ( \{ I_r, H_r, 0\leq r\leq t \} )$ which represent the information on past infections and the combined information on both past infections and hospitalizations, respectively. In line with the original TSI models, we assume the number of new infections at time $t$, given the number of previously infected individuals, follows a Poisson distribution:
\begin{equation}
	I_t \mid \mathcal{F}_{t-1}  \; \sim \; {\rm Poisson}( R_t \Lambda_{t}),  
	\label{countdistribution}
\end{equation}
Next, we introduce a new set of parameters, the hospitalization propensity of an infected individual, $ \tilde{ \omega }_s$, which describes the tendency that an infected individual will be hospitalized $s$ days after infection, independent of the calendar time $t$. For each infected individual, $\tilde{ \omega }_s$ represents the probability that hospitalization occurs $s$ days after infection. For the entire infected population, it represents the proportion of individuals hospitalized $s$ days post-infection. Similar to \( \{\omega_s\}\), the set \( \{\tilde{\omega}_s\} \) satisfies the conditions \( \tilde{\omega}_s \geq 0 \) and \( \sum_{s} \tilde{\omega}_s = 1 \).

We denote $h_{t,s}$ as the number of patients infected at calendar time $t$ and admitted at calendar time $t+s$, i.e., time $s$ since infection, with $\mathbb{E}(h_{t,s}| \mathcal{F}_{t})= \tilde{ \omega }_s I_{t}$. If hospitalization does not occur within a certain period $\tilde{ \eta}$, we assume $h_{t,s}=0$ and $\tilde{ \omega }_s =0$ for $s> \tilde{ \eta}$ for all $t$, where the positive integer $\tilde{ \eta }$ represents the duration from infection to the time after which the likelihood of hospitalization becomes negligible. 
Patients infected at time $t$ and never hospitalized are denoted by $h_{t,-1}$ with the tendency $\tilde{ \omega }_{-1}=1-\sum_{s=0}^{\tilde{ \eta }} \tilde{ \omega }_s$. Thus, total incidence $I_t$ and total hospitalizations $H_t$ at time $t$ are connected through 
\begin{equation}
	I_t = h_{t,-1}+ \sum_{s\geq 0} h_{t,s}, \;\; {\text {and}}\;\; H_t =\sum_{s=0}^{t} h_{t-s,s}, \;\; {\text {for}}\;\; t=0,1,2,\cdots. \label{HospitalAdmissionStructure}
\end{equation}
Based on the above setup, we further assume  
\begin{equation}
	\big(h_{t,-1},h_{t,0},\cdots,h_{t,\tilde{\eta}} \big) \mid I_t \; \sim \; {\rm Multinomial} \big(I_t, \tilde{ \omega }_{-1},  \tilde{ \omega }_0, \cdots, \tilde{ \omega }_{\tilde{ \eta} } \big) \label{MultinormalCondition}.
\end{equation}

\begin{figure}[htp!]
	\begin{center}
		\includegraphics[width=1\textwidth]{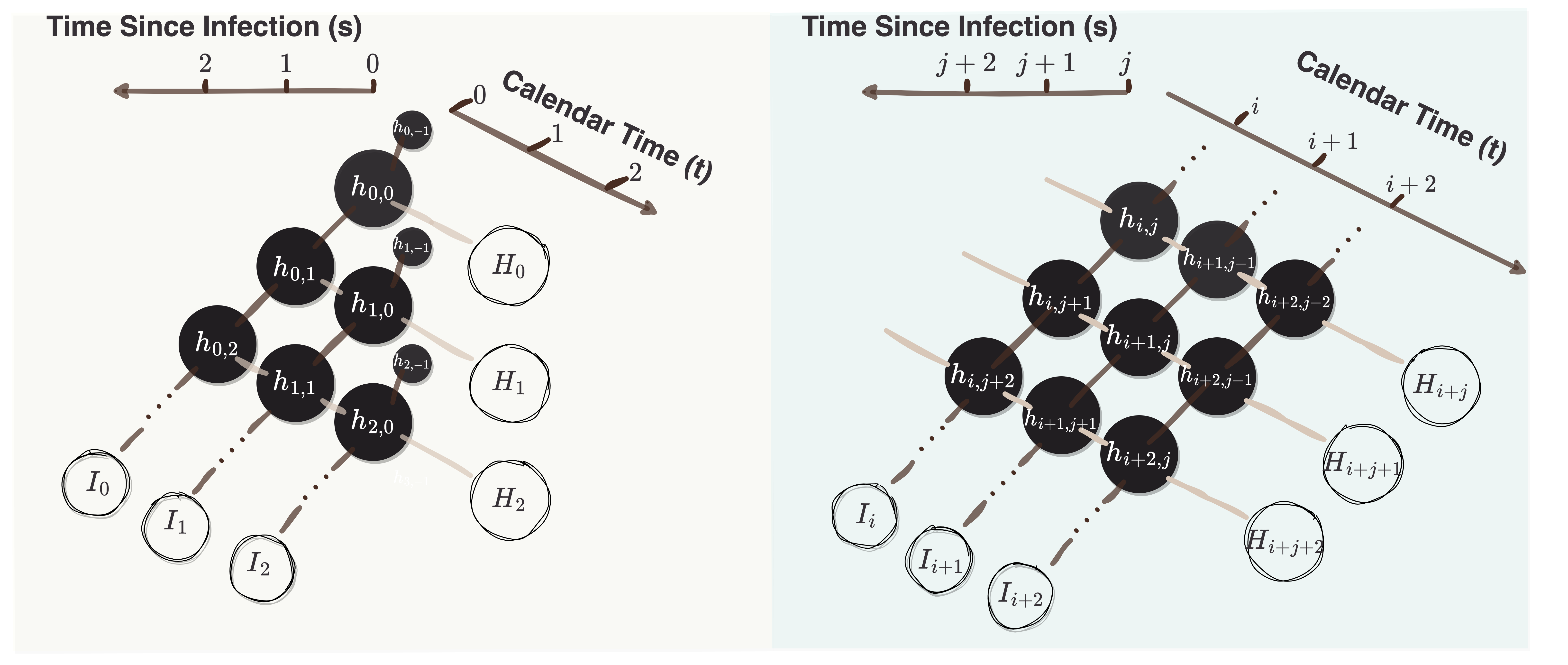}
	\caption{The relationship between incidence data $I_t$ and hospitalization data $H_t$. The left panel illustrates the relationship between calendar time $0$ and $2$ and time since infection from $0$ to $2$. The right panel illustrates the relationship between calendar time $i$ and $i+2$ and time since infection from $j$ to $j+2$.}
	\label{fig:relation}
	\end{center}
\end{figure}

From (\ref{HospitalAdmissionStructure}) and (\ref{MultinormalCondition}), we can see the time-series infection and hospitalization data are deeply interconnected and overlapping (\ref{fig:relation}). To model $R_t$ in relation to risk factors, we assume a regression structure inspired by \cite{shi2022robust} and \cite{quick2021}:
\begin{equation}
	f_{link}(R_t) = X_t^T \beta + \sum_{i=1}^q \theta_i f_i (D_{1,t},D_{2,t},D_{3,t}),  \label{model_assumption} 
\end{equation} 
where $t\geq q>0$ and $X_t$ is a p-dimensional vector that includes a constant 1 and a vector of risk factors of disease transmission, $Z_t$, such as temperature, social distancing measures, and population density. Here, $\beta$ represents the effect size of these exogenous variables on transmission. The link function $f_{link}(\cdot)$ transforms $R_t$, allowing it to be expressed as a linear combination of predictors, e.g., the log link function is one of the most commonly used link functions for non-negative outcomes, while $f_i(\cdot)$ represent a functions of past outcomes. Additionally, we define
\begin{equation}
	D_{1,t} =\big\{ X_r    \big\}_{0\leq r\leq t},\quad D_{2,t}=\big\{ I_r, \mathbb{E} (H_r) \big\}_{0\leq r\leq t-1},\quad D_{3,t}=\big\{ R_r  \big\}_{0\leq r\leq t-1}.  \label{factor}
\end{equation}
With $D_{1,t}$ being observed, equation (\ref{model_assumption}) ensures that $R_t\in \mathcal{F}_t$. Therefore, the second term in (\ref{model_assumption}) captures the temporal structure of disease transmission, while the $\theta_i$'s characterize the level of temporal dependency and continuity in the time series of $\{R_t\}_{t\geq 0}$. By combining  (\ref{countdistribution})-(\ref{factor}), we build a TSI model for hospitalization and incidence data. Indeed, the terms $\{\mathbb{E} (H_r) \}_{1\leq r\leq t-1}$ in $D_{2,t}$ can be expressed as functions of the hospitalization propensity and the $R_t$ according to our model. This implies $D_{2,t} \subset \sigma \Big(  \{  I_r, R_r  \}_{0\leq r\leq t-1}   \Big)$ and hospitalization data are linked with the disease transmission through the hospitalization propensity. In a broader context, one could substitute$D_{2,t}$ in (\ref{factor}) with $\big\{ I_r,  H_r \big\}_{0\leq r\leq t-1}$. We have reserved this general setting to the Supplementary Materials, focusing here on the model shown in Equation (\ref{factor}).

\vspace{-4mm}
\subsection{Composite Likelihood Function}
\label{composite}

Specifying the full likelihood function for the proposed model is challenging. Specifically, the joint distribution of ($I_r, H_r$) given $\mathcal{G}_{r-1}$ for $0\leq r\leq t$ is difficult to retrieve due to the convolution structure outlined in Equation (\ref{HospitalAdmissionStructure}), the irregular boundaries and range restrictions applied to $\{ I_r,H_r \}_{0\leq r\leq t}$, and the fact that $h_{t,s}$ are often unobserved in practice. Therefore, we adopt a composite likelihood approach to estimate the model parameters. Specifically, we use joint distribution of ($I_r, H_r$) given $\mathcal{F}_{r-1}$ instead of  $\mathcal{G}_{r-1}$ to construct a composite log-likelihood function  
\vspace{-4mm}
\begin{equation}
	\ell_{C}=\sum_{0\leq r\leq t} \log  \mathbb{P} \big(H_r, I_r \mid \mathcal{F}_{r-1} \big). \label{composite-likelihood}
\end{equation}
\vspace{-3mm}
We derive (\ref{composite-likelihood}) by summing over all possible values of $\{h_{r-s,s}\}_{1\leq s\leq\tilde\eta}$ in the joint distribution of $(H_r,I_r)$ and  $\{h_{r-s,s}\}_{1\leq s\leq\tilde\eta}$ conditioning on $\mathcal{F}_{r-1}$. To achieve this, we first demonstrate in the following lemma that $h_{t,s}$ given $\mathcal{F}_{t-1}$ follows a Poisson distribution and that $h_{t,{u_1}}$ and $h_{t,{u_2}}$ are independent given $\mathcal{F}_{t-1}$ when $u_1\neq u_2$. Then we derive the form of individual component $\mathbb{P} \big(H_r, I_r \mid \mathcal{F}_{r-1} \big)$ based on this lemma.
\begin{newlemma}
	For arbitrary $t>0$, $-1 \leq u_1,u_2 \leq \min\{ t,\tilde{ \eta} \} $ and $u_1\neq u_2$, we have 
	\begin{equation}
		h_{t,u_1}\mid  \mathcal{F}_{t-1} \sim Poisson ( \tilde{ \omega }_{u_1} R_{t} \Lambda_t),  \;\; \;\; h_{t,u_2}   \mid  \mathcal{F}_{t-1} \sim Poisson (\tilde{ \omega }_{u_2} R_{t} \Lambda_t) ,
	\end{equation}
	and $h_{t,u_1} \mid \mathcal{F}_{t-1} \perp h_{t,u_2} \mid \mathcal{F}_{t-1}$. Moreover, $(h_{t,u_1}, h_{t,u_2} ) \mid \mathcal{G}_{t-1}=(h_{t,u_1}, h_{t,u_2} ) \mid \mathcal{F}_{t-1}$, and  $h_{t,u_1} \mid \mathcal{G}_{t-1} \perp h_{t,u_2} \mid \mathcal{G}_{t-1}$.
	\label{PoissonDistribution}
\end{newlemma}
According to  Equation (\ref{MultinormalCondition}) and Lemma \ref{PoissonDistribution}, if we use $\mathbbm{1}(\cdot)$ to denote the indicator function, which equals 1 when its condition is true and 0 otherwise, the joint distribution of $(H_r,I_r)$ and  $\{h_{r-s,s}\}_{1\leq s\leq\tilde\eta}$ conditioning on $\mathcal{F}_{r-1}$ can be expressed as the product of the probability density functions of several binomial distributions and two Poisson distributions as follows:
\begin{align}	\label{dist_with_missing}
	&\mathbb{P}\big(H_r, I_r,h_{r-\tilde{\eta},\tilde{ \eta}}, \cdots, h_{r-1,1} \mid \mathcal{F}_{r-1}\big)\\
	=&\mathbb{P}\bigg( h_{r-\tilde{\eta},\tilde{ \eta}}, \cdots, h_{r-1,1},  h_{r,0}=H_r-\sum_{s=1}^{\tilde{ \eta}} h_{r-s,s} ,  \sum_{s=-1,1,\cdots,\tilde\eta}  h_{r,s}=I_r-H_r+\sum_{s=1}^{\tilde{ \eta}} h_{r-s,s}	\Big| \mathcal{F}_{r-1}\bigg) \nonumber\\
%	&\mathbb{P}\big(H_r, I_r,h_{r-\tilde{\eta},\tilde{ \eta}}, \cdots, h_{r-1,1} \mid \mathcal{F}_{r-1}\big)\\
	= & \prod_{s=1}^{\tilde{ \eta}} \mathbb{P} \Big( Binomial( I_{r-s},\tilde{ \omega }_s ) = h_{r-s,s} \Big)   \mathbbm{1} \Big(  H_r-I_r	\leq  \sum_{s=1}^{\tilde{ \eta}} h_{r-s,s} \leq H_r  \Big)    \nonumber\\
	&  \qquad \;\;      \cdot \mathbb{P}\Big( Poisson(\tilde{ \omega }_0 R_{r} \Lambda_r ) = H_r-\sum_{s=1}^{\tilde{ \eta}} h_{r-s,s} \Big)  \nonumber\\
	&  \qquad   \;\; 	 \cdot  \mathbb{P}\Big( Poisson((1-\tilde{ \omega }_0) R_{r} \Lambda_r ) = I_r-H_r+\sum_{s=1}^{\tilde{ \eta}} h_{r-s,s} \Big)\nonumber
\end{align}

Therefore, the joint distribution of $(H_r,I_r)$ conditioning on $\mathcal{F}_{r-1}$ can be calculated by summing over all possible values of $\{h_{r-s,s}\}_{1\leq s \leq \tilde\eta}$,
\begin{equation}
	\mathbb{P} \big(H_r, I_r  \mid \mathcal{F}_{r-1} \big)  = \sum_{h_{r-\tilde{\eta},\tilde{ \eta}}, \cdots, h_{r-1,1}} \mathbb{P}\big(H_r, I_r,h_{r-\tilde{\eta},\tilde{ \eta}}, \cdots, h_{r-1,1} \mid \mathcal{F}_{r-1}\big). \label{dist}
\end{equation}

\section{Inference}

\subsection{Estimation}
\label{MCEM}
%The calculation of the composite log-likelihood function involves summing over all possible values of  $\{ h_{r-s,s}, 1\leq s\leq \tilde{ \eta } \}$. 
In this section, we describe the procedures for estimating model parameters using MCEM algorithms, where the missing data $\{h_{r-s,s}\}_{1\leq s \leq \tilde\eta}$ are unobserved tracing information and the observed disease data consist of daily incidence and hospitalization counts. Let $\gamma=(\beta,\theta,\omega,\tilde{ \omega })\in {\bf\Gamma}$ denote the model parameters, where $\beta$ and $\theta$ define a time series model for the instantaneous reproduction number $\{ R_t \}_{t\geq 1}$, as described in equation (2.4). The terms  $\{ R_t \}_{t\geq 1}$ are integrated into the composite likelihood function (\ref{composite-likelihood}) through (\ref{dist_with_missing}) and (\ref{dist}), introducing  $\beta$ and $\theta$ as parameters to be estimated. We use notations like $\mathbb{P}_{\gamma}$ to indicate that the probability is associated with the parameter value $\gamma$. We use $\gamma_0$ to denote the parameter values corresponding to the underlying true data-generating mechanism, and $\hat{\gamma}$ denote the maximum composite likelihood estimator. The composite likelihood function integrates information from both the observed data and the modeled structure of $\{ R_t \}_{t\geq 1}$, and the EM algorithm addresses the missing data, enabling the joint estimation of all parameters $\gamma$. To improve the efficiency of sampling the missing data $h_{r-\tilde{\eta},\tilde{\eta}},\cdots, h_{r-1,1}$ in the algorithm, we incorporate an acceptance-rejection sampling method.

Let $Data_{obs, r} = \{ X_j, I_j  \}_{0\leq j\leq r}  \cup \{ H_r \}$ denote the observed data and $Data_{miss,r}= \{ h_{r-s,s}  \}_{1\leq s\leq \tilde{ \eta}}$ denote the missing data at time $r$. In the MCEM algorithm, $N_0$ denotes the Monte-Carlo sample size, and $Data_{miss,r}^{(m,k)}$, $1\leq m\leq N_0$ denotes the $m$-th Monte-Carlo sample in the $k$-th iteration of the EM algorithm. Throughout the following, $(k)$ denotes the $k$-th EM iteration. For example, $\gamma^{(k)}$ represents the current estimate of $\gamma$ after the $k$-th EM iteration, with $R_r^{(k)}$ and $\Lambda_r^{(k)}$ representing estimates of $R_r$ and $\Lambda_r$ based on (\ref{countdistribution}) and (\ref{model_assumption}), using parameter values $\gamma^{(k)}$. The estimation procedure is outlined below, with a pseudo-code provided in Algorithm \ref{MCEM1}:
\vspace{-1mm}
\begin{itemize}
	\item[\bf E step] At the $(k+1)$-th iteration, given the current estimate $\gamma^{(k)} = (\beta^{(k)},\theta^{(k)},\omega^{(k)},\tilde{ \omega }^{(k)} )$, this step computes the expected complete data composite log-likelihood, or the  Q-function:
		\vspace{-2mm}
	\begin{equation}
		Q( \gamma \mid \gamma^{(k)}  ) \overset{\rm def}{=}  \sum_{0\leq r\leq t}  \mathbb{E}_{\gamma^{(k)}}   \Big( \log \mathbb{P}_{\gamma} \big(H_r, I_r, Data_{miss,r} \mid \mathcal{F}_{r-1} \big)  \mid Data_{obs,r}  \Big), \label{Qfunction}
		\vspace{-3mm}
	\end{equation}	
	where the expectation is taken over the distribution of $Data_{miss,r}$ conditional on $( Data_{obs,r}, \gamma^{(k)})$. Due to computational complexity, the Q-function is approximated using Monte Carlo sampling of $Data_{miss,r}$ from $\mathbb{P}_{\gamma^{(k)}} \big(Data_{miss,r} \mid Data_{obs,r} \big)$, and is calculated as
	\begin{equation*}
		\hat{Q}( \gamma \mid \gamma^{(k)}  ) \overset{\rm def}{=}  \sum_{0\leq r\leq t} \frac{1}{N_0}\sum_{m=1}^{N_0} \log \mathbb{P}_{\gamma} \big(H_r, I_r, Data_{miss,r}^{(m,k)} \mid \mathcal{F}_{r-1} \big).
		\vspace{-3mm}
	\end{equation*}	
	\item[\bf M step]
	This step is to compute $\gamma^{(k+1)} = \arg\max_{\gamma \in \Gamma} \hat{Q}(\gamma \mid \gamma^{(k)})$. If $\arg\max_{\gamma \in \Gamma} \hat{Q}(\gamma \mid \gamma^{(k)})$ is not unique, we  randomly choose one as $\gamma^{(k+1)}$. When $\hat Q(\gamma^{(k)}\mid \gamma^{(k)})=\max_{\gamma \in \Gamma} \hat{Q}(\gamma\mid \gamma^{(k)})$, we choose $\gamma^{(k+1)}=\gamma^{(k)}$.
\end{itemize}

Notably, drawing samples from \(P_{\gamma^{(k)}} \left(Data_{\text{miss},r} \mid Data_{\text{obs},r} \right)\) is equivalent to sampling from \(P_{\gamma^{(k)}} \left(H_r, I_r, Data_{\text{miss},r} \mid \mathcal{F}_{r-1} \right)\), as $P_{\gamma^{(k)}}(H_r,I_r \vert \mathcal{F}_{r-1})$ in equation \eqref{dist} is constant for a given $\gamma^{(k)}$. To improve sampling efficiency, we use an acceptance-rejection method in Algorithm \ref{MCEM1}. Specifically, samples are first drawn from  \(P_{\gamma^{(k)}} \left(Data_{\text{miss},r} \mid \mathcal{F}_{r-1} \right)\) and accepted with probability \(p_{\text{acceptance}}\), proportional to \(P_{\gamma^{(k)}} \left(H_r, I_r \mid Data_{\text{miss},r}, \mathcal{F}_{r-1} \right)\), where $\max_{H_r,I_r}p_{\text{acceptance}}\leq 1$. To further enhance efficiency, particularly when infections and hospitalizations are large ($H_t \geq 25$) in later pandemic stages, \(p_{\text{acceptance}}\) in Algorithm \ref{MCEM1} can be adjusted using Stirling's approximation:
\begin{equation*}
	p_{\text{acceptance-adj}} = 2\pi R_r^{(k)} \Lambda_r^{(k)} \sqrt{\tilde{\omega}_0^{(k)}  ( 1 - \tilde{\omega}_0^{(k)} )} p_{\text{acceptance}}.
	\vspace{-2mm}
\end{equation*}
%This adjustment improves computational speed in practice.

\begin{algorithm}
\caption{An MCEM algorithm to estimate model parameters with unknown $\omega_s$ and $\tilde\omega_s$.}
	\label{MCEM1}
	\begin{algorithmic}
		\Require initial parameter value $\gamma^{(0)}$, $k\leftarrow0$, breakpoint critical value $\Delta_0$.
		\While {$\Vert \gamma^{(k+1)} -\gamma^{(k)} \Vert_{\infty} > \Delta_0  $, \textbf{set} $r\leftarrow0$, $m\leftarrow1$,}
		\While {$0\leq r\leq t$, $1\leq m \leq N_0$,}
		\State Sample  $h_{r-s,s}$ independently from Binomial$(I_{r-s}, \tilde{ \omega }_s^{(k)})$, for $1\leq s\leq  \tilde{ \eta }$.
		\State Sample $\psi$ from Bernoulli distribution with probability $p_{\text{acceptance}}$, where	\State $p_{\text{acceptance}}=\mathbb{P}\Big( Poisson(\tilde{ \omega }_0^{(k)} R_{r}^{(k)} \Lambda_r^{(k)} ) = H_r-\sum_{s=1}^{\tilde{ \eta}} h_{r-s,s} \Big)$\State $\times \mathbb{P} \Big( Poisson((1-\tilde{ \omega }_0^{(k)}) R_{r}^{(k)} \Lambda_r^{(k)} ) = I_r-H_r+\sum_{s=1}^{\tilde{ \eta}} h_{r-s,s} \Big)$
		\If {$\psi=1$,}
		\State let $Data_{miss,r}^{(m,k)}=\{h_{r-s,s}\}_{1\leq s\leq \tilde{ \eta }}$, $m\leftarrow m+1$.
		\EndIf 
		\If {$m>N_0$,} 
		\State let $r\leftarrow r+1$, $m\leftarrow1$.
		\EndIf
		\EndWhile 
		\State Calculate the Monte Carlo Q-function $\hat{Q}(\gamma \mid \gamma^{(k)})$ and $\gamma^{(k+1)} = \arg\max_{\gamma \in \Gamma} \hat{Q}(\gamma \mid \gamma^{(k)})$.
		\State Let $k\leftarrow k+1$, $\hat{\gamma}\leftarrow \gamma^{(k+1)}$.
		\EndWhile
		\State Output $\hat{\gamma}$.
	\end{algorithmic}
\end{algorithm}

In Algorithm \ref{MCEM1}, the parameters of the infectiousness function, $\omega_s$, and hospitalization propensity, $\tilde\omega_s$, are assumed to be unknown. This scenario is often encountered in the early stages of novel pandemics when little is known about the infectious pathogens. However, prior knowledge about $\omega_s$ and $\tilde\omega_s$ may sometimes be available from multiple biomedical or contact tracing studies. In such scenarios, we may consider these estimates from other studies as prior knowledge. In the Supplementary Materials, we provide another MCEM algorithm, Algorithm 2, to estimate the model parameters when prior knowledge about $\omega_s$ and $\tilde\omega_s$ is available.

\subsection{Asymptotic Properties}
\label{asymp}
The basic TSI model (\ref{countdistribution}), which describes the generative nature of an infectious pathogen, shares similarities with a branching process. Since branching process degenerates on an extinction set with no asymptotic properties, we similarly define an extinction set $\mathcal{E}$ for the TSI model,
\begin{equation}
	\mathcal{E} \overset{\rm def}{=} \Big\{ I_r=0,\; {\rm for}\;  r {\rm \ greater \  than \ some \ } K  \Big\} = \bigcup_{r\geq 1} \Big\{ \frac{\partial \log \mathbb{P} (I_r,H_r \mid \mathcal{F}_{r-1}) }{\partial \gamma} =0 \Big\},  \label{non-extinction set} 
\end{equation}
and define $\mathcal{E}_{none}=\mathcal{E}^c$. Often, no consistent estimator $\hat{\gamma}$ exists in $\mathcal{E}$. Thus, we set aside the extinction probability 
\begin{equation*}
	\mathbb{P}\big(  \mathcal{E}   \big) = \mathbb{P}\Big(  \bigcup_{r\geq 1} \Big\{ \frac{\partial \log \mathbb{P} (I_r,H_r \mid \mathcal{F}_{r-1}) }{\partial \gamma} =0 \Big\}   \Big)
\end{equation*}
and focus on the asymptotic behavior of $\hat{\gamma}$ on the non-extinction set $\mathcal{E}_{none}$. 

In the following theorems, we show that the proposed MCEM algorithm preserves the ascent property of the composite likelihood function and converges for our model. We then establish the consistency of the maximum composite likelihood estimator and present it in the practical form of Equation (\ref{model_assumption}), under certain regularity conditions. Proof details are provided in the Supplementary Materials.
\begin{cctheorem}[Ascent property of the composite likelihood]
	\label{AscentProperty}
	For $k \geq 0$, we have $\ell_{C}( \gamma^{(k+1)} ) \geq \ell_{C} ( \gamma^{(k)})$.  
\end{cctheorem}
\begin{cctheorem}[Convergence of the MCEM algorithm]
	\label{StationaryConvergence}
	Assume $\Gamma$ is a compact set, then with $k\rightarrow\infty$, the MCEM-estimator $\gamma^{(k)}$ at the $k$-th iteration converges to one of the stationary point $\gamma_s$ induced by $M(\cdot)=\arg\max_{\gamma \in \Gamma} Q( \gamma \mid \cdot )$, and $\ell_C(\gamma^{(k)})$ converges monotonically to $\ell_C(\gamma_s)$.
\end{cctheorem}

To establish the consistency of the maximum composite likelihood estimator, we impose regularity conditions on the observed time series. Inspired by work on counting processes \citep{zeger1988markov, davis1999modeling, davis2000autocorrelation, davis2003observation}, we assume:
\begin{condition}[Series ergodicity]\label{condition of consistency}
	There exist $t_0>0$, such that,
	\begin{equation*}
		\lim_{t\rightarrow\infty} \frac{t_0}{t} \sum_{s=1}^t R_s\Lambda_{s} \rightarrow_{a.s.} \sum_{s=1}^{t_0} \mathbb{E} R_s\Lambda_{s}, \quad \lim_{t\rightarrow\infty} \frac{t_0}{t}  \sum_{s=1}^t I_s \log (R_s) \rightarrow_{a.s.} \sum_{s=1}^{t_0} \mathbb{E} I_s \log (R_s). 
		%\quad \lim_{t\rightarrow\infty} \f{t_0}{t}  \sum_{s=1}^t \log R_s = \sum_{s=1}^{t_0} \e \log R_s.
	\end{equation*}
\end{condition}

We also assume the time series regression in Equation (\ref{model_assumption}) uses a log link function and follows an autoregressive model of order 1, such that $\log(R_t) = Z_t^T \beta + \theta_0+ \theta_1 \log(R_{t-1})$, where $\theta_0$ is an intercept term extracted from both the exogenous terms and the autoregressive terms and \(\theta_1\) serves as the autoregressive parameter, capturing the dependence of \( R_t \) on its previous value. We then establish the consistency of the maximum composite likelihood estimator for the time series regression coefficients when the infectiousness function and hospitalization propensity,  $(\omega_s, \tilde\omega_s )$, are known. We denote the parameters of interest as $\gamma\vert_{\omega}=(\beta,\theta)$. The consistency of the maximum composite likelihood estimator,  $\hat{\gamma}\vert_{\omega}=(\hat{\beta},\hat{\theta})$, is shown as follows.

\begin{cctheorem}\label{consistency}
	(Strong consistency of the estimators) Under condition \ref{condition of consistency} and assuming that $\Gamma$ is a compact set, we have $\hat{\gamma}\vert_{\omega} \xrightarrow[]{a.s.}\gamma_0\vert_{\omega}$ as the length of the observational days $t\rightarrow \infty$,
	where $\gamma_0\vert_{\omega}$ is the parameter value corresponding to the true data generating mechanism. 
\end{cctheorem} 
Furthermore, with the following conditions from the theorem 3 of \cite{kaufmann1987regression}, that is
\vspace{-3mm}
\begin{condition}[Non-singularity]\label{condition of normality1}
	For each $1\leq r\leq t$, define
	\begin{equation*}
		\xi_r (\gamma\vert_{\omega}) = \frac{ \partial R_r}{ \partial \gamma\vert_{\omega} } \cdot R_r^{-1} \cdot \big( I_r - R_r \Lambda_r     \big).
	\end{equation*}
	There exists some nonrandom and non-singular normalizing matrix $A_t$, such that the normalized conditional variance converges to an almost surely positive definite random matrix $\zeta^T\zeta$, i.e., 
	\begin{equation*}	 
		A_t^{-1} \Big[ \sum_{r=1}^{t}\text{Cov} \big(   \xi_r (\gamma_0\vert_{\omega}) \mid \mathcal{F}_{r-1}    \big)	  \Big] \big( A_t^{-1} \big)^T \overset{P}{\longrightarrow} \zeta^T \zeta.
	\end{equation*}
\end{condition}

\begin{condition}[Uniformly integrability]\label{condition of normality4}
	For $1\leq r\leq t$, $\mathbb{E} \big[ \xi_r^2(\gamma_0) \mid \mathcal{F}_{r-1} \big]$ is termwise uniformly integrable.
\end{condition}
\vspace{-8mm}
\begin{condition}[The conditional Lindeberg condition]\label{condition of normality2}
	For arbitrary $\epsilon>0$,			
	\begin{equation*}
		\sum_{r=1}^t \mathbb{E} \Big[ \xi_r^T(\gamma_0\vert_{\omega}) \big(A_t^T A_t\big)^{-1} \xi_r(\gamma_0\vert_{\omega}) \cdot \mathbbm{1}(|\xi_r^T(\gamma_0\vert_{\omega})  \big(A_t^T A_t\big)^{-1} \xi_r(\gamma_0\vert_{\omega})|>\epsilon^2) \big| \mathcal{F}_{r-1}  \Big] \overset{P}{\longrightarrow} 0.
	\end{equation*}
\end{condition}

\begin{condition}[The smoothness condition]\label{condition of normality3}
	For arbitrary $\delta>0$, and $\delta-$neighborhood ball $\mathcal{B}_t (\delta)$ defined as $\mathcal{B}_t (\delta)= \{ \tilde{\gamma}\vert_{\omega}:  \Vert A_t^T (\tilde{\gamma}\vert_{\omega} - \gamma_0\vert_{\omega}) \Vert \leq \delta  \}$,
	\begin{equation*}
		\sup_{\tilde{\gamma}\vert_{\omega}\in \mathcal{B}_t (\delta)} \bigg \Vert  A_t^{-1} \sum_{r=1}^{t} \Big(  \frac{\partial \xi_r(\tilde{\gamma}\vert_{\omega} )}{\partial \gamma}	+ \text{Cov}  \big(   \xi_r(\gamma_0\vert_{\omega}) \mid \mathcal{F}_{r-1}    \big)	  \Big)   \big( A_t^{-1} \big)^T     \bigg \Vert \overset{P}{\longrightarrow} 0.
	\end{equation*}
\end{condition}
\noindent the maximum composite likelihood estimator converges to a normal distribution as shown below.
\vspace{-10mm}
\begin{cctheorem}[Asymptotic normality]\label{normality}
Under condition \ref{condition of consistency}-\ref{condition of normality3}, and assume $\Gamma$ is a compact set, then on the non-extinction set defined in (\ref{non-extinction set}), with $t\rightarrow \infty$, 
	\begin{equation}	
		\vspace{-5mm} 
		\Big[ \sum_{r=1}^{t}\text{Cov}  \left(   \xi_r (\gamma_0\vert_{\omega}) \mid \mathcal{F}_{r-1}    \right)	  \Big]^{1/2} \big(\hat{\gamma}\vert_{\omega}-\gamma_0\vert_{\omega} \big) \overset{d}{\longrightarrow} N(0,I), \label{normal}
	\end{equation}
\end{cctheorem}
\noindent where $I$ is the identity matrix.

\section{Simulation Studies}
\label{simulation}
\subsection{Simulation Setups}
To evaluate the performance of the proposed method, we conducted simulation studies under two sets of scenarios. In the first, reported new infections represent the true counts, meaning the TSI model is correctly specified (correctly specified model). In the second, reported counts include errors (e.g., under-reporting or limited testing), implying the TSI model is misspecified (misspecified model). In both scenarios, hospitalization data were assumed accurate. Each scenario was repeated 1,000 times to assess bias and coverage probability of the estimator.

Daily infections, hospitalizations, and covariates were generated from the proposed model (\ref{countdistribution})-(\ref{factor}), with a simplified form of (\ref{model_assumption}):
%\begin{equation}
$\log(R_{r})=Z_{r}^T\beta+\theta_0+\theta_1\log(R_{r-1}), $ for $r\geq 1
%\label{eq:generateRt}
$, which means that the logarithm of the instantaneous reproduction number $\{ R_r\}_{1\leq r\leq t}$ follows an AR$(1)$ structure with exogenous terms. Following \cite{li2020early}, the infectiousness function, approximated by the serial interval, was reported to follow a Gamma distribution with a mean of 7.5 days (95\% CI: [5.3, 19]), corresponding to a standard deviation ranging from approximately 1.12 to 5.86. Our simulations used a Gamma distribution for the infectiousness function, with shape and scale parameters adjusted to mirror these results. Similarly, hospitalization propensity was set to a Gamma distribution to mimic the delay between disease onset and hospitalization reported in the same study:
\vspace{-3mm}
\begin{gather*}
	\omega_s= \mathbb{P} \big(  \Gamma(k_1,\mu_1) \in [s-1,s)     \big), \; \; s=1,\cdots, 24, \quad {\rm and}\;\; \omega_{25}=\mathbb{P} \big( \Gamma(k_1,\mu_1) \geq 24     \big), \\
	2\tilde{\omega}_s= \mathbb{P} \big(  \Gamma(k_2,\mu_2) \in [s, s+1)     \big), \; s=0,\cdots,4, \; 2\tilde{\omega}_5= \mathbb{P} \big(  \Gamma(k_2,\mu_2) \geq 5  \big), \;\; {\rm and}\; \tilde{\omega}_{-1}=0.5,
\end{gather*}
\noindent where $\Gamma(\cdot,\cdot)$ had shape parameters $\check{k}_1=2.5$, $\check{k}_2=1.6$, and scale parameters $\mu_1=3$, $\mu_2=1.5$. We set the study duration $T=120$ and fixed $p=2$. Parameter values were assigned as $(\theta_0,\theta_1,\beta^T)=(0.7,0.5,-.02,-.125)$.  We independently simulated two =covariates $\{Z_{r,1},Z_{r,2}\}_{1\leq r\leq t}$ to mimic real data on temperature in Philadelphia and social distancing trends obtained from daily cellular telephone movements, as provided by Unacast (\cite{unacast}, July 1st). This data represented the percentage change in visits to non-essential businesses, such as restaurants and hair salons, between March 1st and June 30th, 2020.

\vspace{-5mm}

\subsection{Simulation Results with Correctly Specified Model}
\begin{figure}[htp!]
	\begin{center}
		\includegraphics[width=1\textwidth]{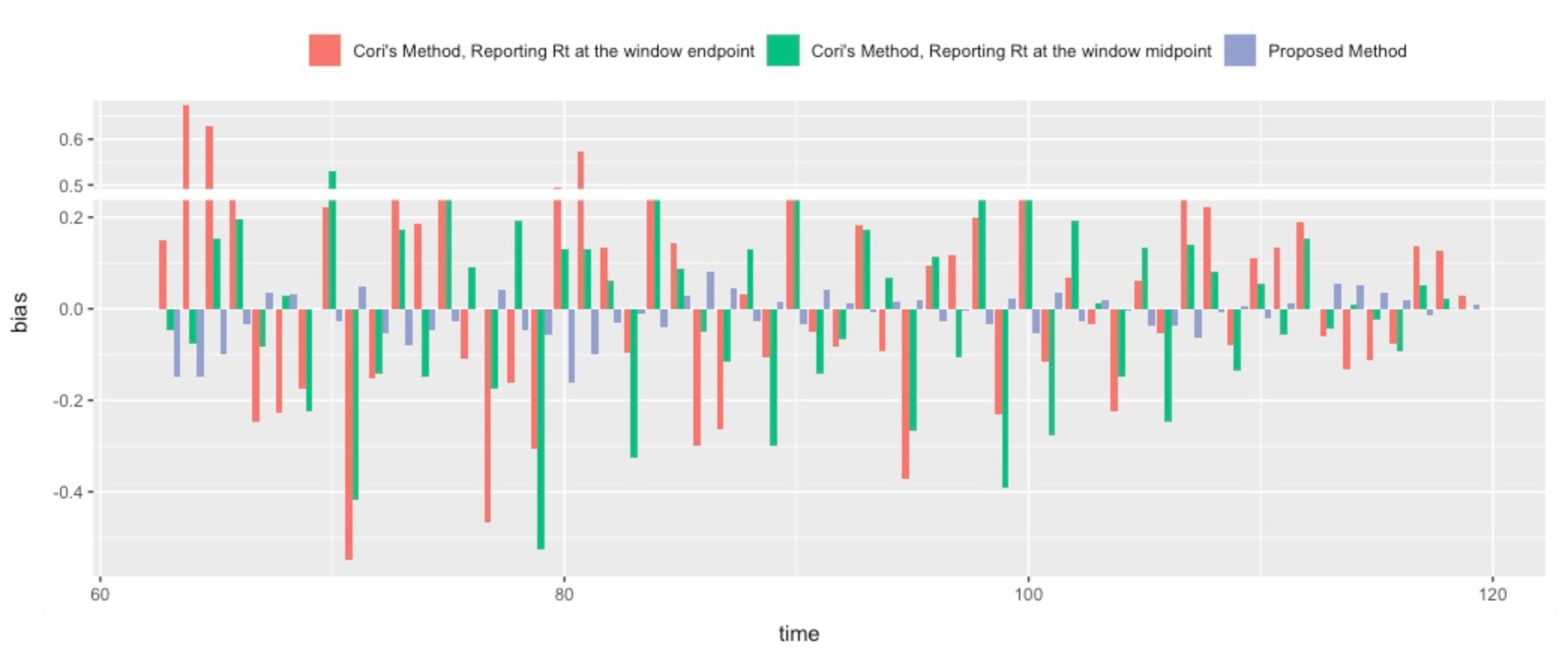}
		\caption{Comparison of the estimation bias in the daily instantaneous reproduction number, $R_t$, between the proposed method, Cori's method that reports $R_t$ at the endpoint of the sliding window \citep{cori2013new}, and Cori's method that reports $R_t$ at the midpoint of the sliding window \citep{gostic2020practical,gressani2022epilps}, when the reported number of daily new infections was accurate, and both the infectiousness function $\omega_s$ and hospitalization propensity $\tilde\omega_s$ were known. For the reference approach, a three days sliding window is selected by minimizing the $\mathcal{L}_2$-distance of the estimated and oracle sequence of $\{R_r\}_{1\leq r\leq t}$, and a comparison to reference approach with different sliding window is left to the supplementary. The estimation bias was calculated as the difference between the estimated daily $R_t$ and the oracle $R_t$ that generates the data. The bias is plotted starting from day 60, as \cite{cori2013new} demonstrated substantial bias and unstable estimation during the early stages of the simulated period when the incident cases $I_t$ were small ($\leq 2 \times 10^3$).}%, which makes it difficult to present the bias difference.}
		\label{fig:bias}
	\end{center}
\end{figure}

In this set of scenarios, we explored three circumstances and compared the proposed method to the reference approach by \cite{cori2013new}. In the first circumstance, we assumed the infectiousness function $\omega_s$ and hospitalization propensity $\tilde\omega_s$ were known. The model parameters reduced to $\gamma=(\theta^T, \beta^T)$. For the reference approach, a 3-day sliding window was chosen by minimizing the $\mathcal{L}_2$-distance between the estimated and oracle $\{R_r\}_{1\leq r\leq t}$. Additional results for 1-day and 7-day windows are in the Supplementary Materials. For the proposed method, we selected the initial $\gamma^{(0)}$ randomly and away from the oracle value. Figure \ref{fig:bias} shows both methods captured the trend of $\{R_r\}_{1\leq r\leq t}$ well with very small estimation biases, but the proposed method performed better than the baseline method with a smaller bias. Moreover, when $\omega_s$ and $\tilde\omega_s$ were known, we found the proposed MCEM algorithm converged very fast, and the estimation bias for $\{R_r\}_{1\leq r\leq t}$ reached its limit after only two iterations of running the algorithm. In the second circumstance, we assumed $\omega_s$ and $\tilde\omega_s$ were unknown but prior knowledge about $\omega_s$ was available. We used the estimated infectiousness functions from previous studies as the prior knowledge, \citep{wu2020estimating, ali2020serial, chen2022inferring,deng2021estimation}, set the true function according to the results in \cite{li2020early}, and used Algorithm 2 to estimate the parameters. Results were similar to those observed in the first circumstance, except for a small increase in estimation bias for both methods. In the third circumstance, we assumed $\omega_s$ and $\tilde\omega_s$ were unknown and no prior knowledge was available. In this situation, we only estimated the parameters using the proposed method, since the reference approach either requires user-specified values to constrain the overall shape of the infectiousness function or needs user-provided contact tracking data to estimate the infectiousness function. Using the proposed method, we estimated the regression coefficients as well as the infectiousness function and hospitalization propensity. As shown in Table \ref{estimationResults}, the proposed method produced accurate estimates for parameters with small bias and good coverage probability. It also consistently estimated the instantaneous reproduction numbers (Figure S7 in Supplemental Materials). Overall, when the TSI model was correctly specified, the proposed composite likelihood MCEM algorithm benefited from incorporating hospital admission data, outperforming the reference approach.

\vspace{-5mm}

\subsection{Simulation Results with Misspecified Model}
\label{misspecified}

In the second set of scenarios, we assumed the reported daily new infections were inaccurate due to underreporting, with the proportion of reported cases varying daily. This reflects real-world challenges during the pandemic, where case reporting was influenced by testing availability, public compliance, and at-home testing. In contrast, hospitalization data were assumed accurate, as hospitals report admissions in standardized formats. Additional simulations involving underreporting of hospitalizations are included in the Supplementary Material.

\begin{figure}[!p]
	\begin{center}	\includegraphics[width=1\textwidth]{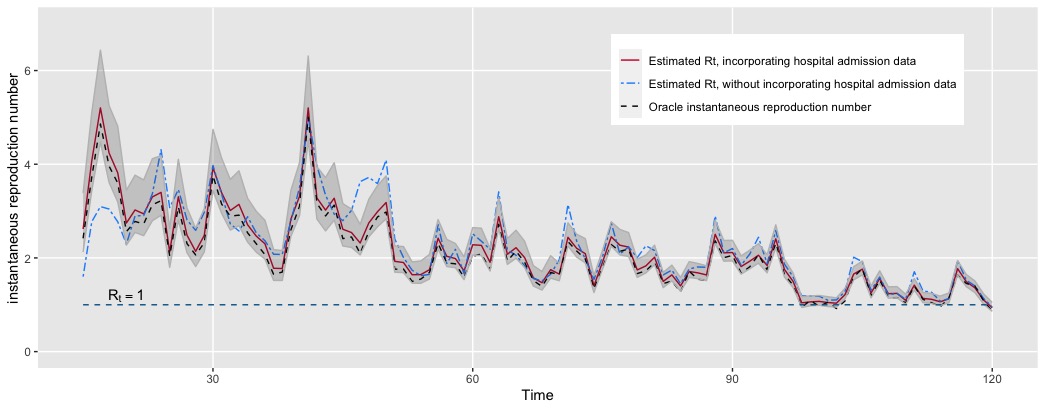}
		\caption{Estimation of the instantaneous reproduction number when the daily new infections were reported with 0\%-30\% under-reporting rates. Black dotted line stands for the oracle instantaneous reproduction number. Red solid line and the gray shadow stand for the estimates and its corresponding Bootstrapping confidence interval ($90\%$ confidence level) using the proposed method. Blue dashed line stands for the estimates using a comparison method \citep{shi2022robust}.}
		\label{fig:comparisons}
	\end{center}
\end{figure}

Daily underreporting percentage was generated from a normal distribution with a mean of 15\% and standard deviation of 5\%. This allowed the daily under-reporting percentage to vary from 0\% to 30\%, resulting in poor data quality for the reported infection numbers. However, we assumed the daily hospital admission data were accurate, ensuring that all hospitalized patients were reported and documented in a timely manner. Using the oracle reproduction number, we simulated true infections, underreporting percentages, and hospitalizations for each replication.
The proposed method produced results similar to the correctly specified model. For example, the mean point estimates for $\beta$ in the mis-specified model (-0.0206, -0.1239) were nearly identical to those in the correctly specified model (-0.02, -0.125), though the standard errors were higher in the mis-specified scenario (0.003, 0.009) compared to the correctly specified model (0.0003, 0.0004). Compared to a recent measurement error model \citep{shi2022robust}, the proposed method performed slightly better, providing more accurate estimates of the reproduction number (Figure \ref{fig:comparisons}).

\vspace{-3mm}
\section{Application to COVID-19 Data}
\label{application}
We applied the proposed method to COVID-19 data from January 1 to June 30, 2021, for four major metropolitan counties: Miami-Dade, FL; New York, NY; Cook (Chicago), IL; and Wayne (Detroit), MI. These counties represent four major metropolitan areas in the Southeast coast, Tri-State area, and Great Lakes region of the United States. County-level data on daily new infections and hospitalizations due to COVID-19 were obtained from the National Healthcare Safety Network (NHSN) database. Additionally, two county-level risk factors were sourced: daily social distancing practices, indicated by the percentage change in visits to nonessential businesses (from Unacast), and wet-bulb temperature (from the National Oceanic and Atmospheric Administration). These risk factors were selected based on a thorough review of the literature that has extensively examined local factors influencing COVID-19 transmission \citep{talic2021effectiveness,weaver2022environmental,rubin2020association}. Our objectives were to estimate the daily reproduction number $R_t$, infectiousness function $\omega_s$, and hospitalization propensity $\tilde\omega_s$, and to assess the association between county-level factors and disease transmission. These findings aim to inform public health policies and resource allocation.

When fitting the proposed model, we allowed hospitalization propensity to vary by county to account for differences in healthcare resources and access. Due to limitations in US COVID-19 surveillance data, which recorded incidence at the time of disease diagnosis rather than actual infection, and considering that the exact infection times are rarely known, we interpreted the infectiousness function and hospitalization propensity estimated from this data based on diagnosis time rather than infection time. Specifically, $\tilde{\omega}{-1,c}$ and $\tilde{\omega}{s,c}$ represent the probabilities of never being hospitalized and being hospitalized on the $s$-th day after diagnosis, respectively, for each county $c\in{1,2,3,4}$. This approach captures county-level heterogeneity, while $\omega_s$ reflects infectiousness on the $s$-th day post-diagnosis. Bootstrap confidence intervals (CIs) were computed using the block approach \citep{buhlmann1999}. The time series structure (\ref{model_assumption}) was selected via AIC from a family of autoregressive models with exogenous terms, and the lengths of the infectiousness function and hospitalization propensity ($\eta$ and $\tilde{\eta}$) were determined by maximizing the composite likelihood.

\begin{figure}[!p]
	\begin{center}
		\includegraphics[width=1\textwidth]{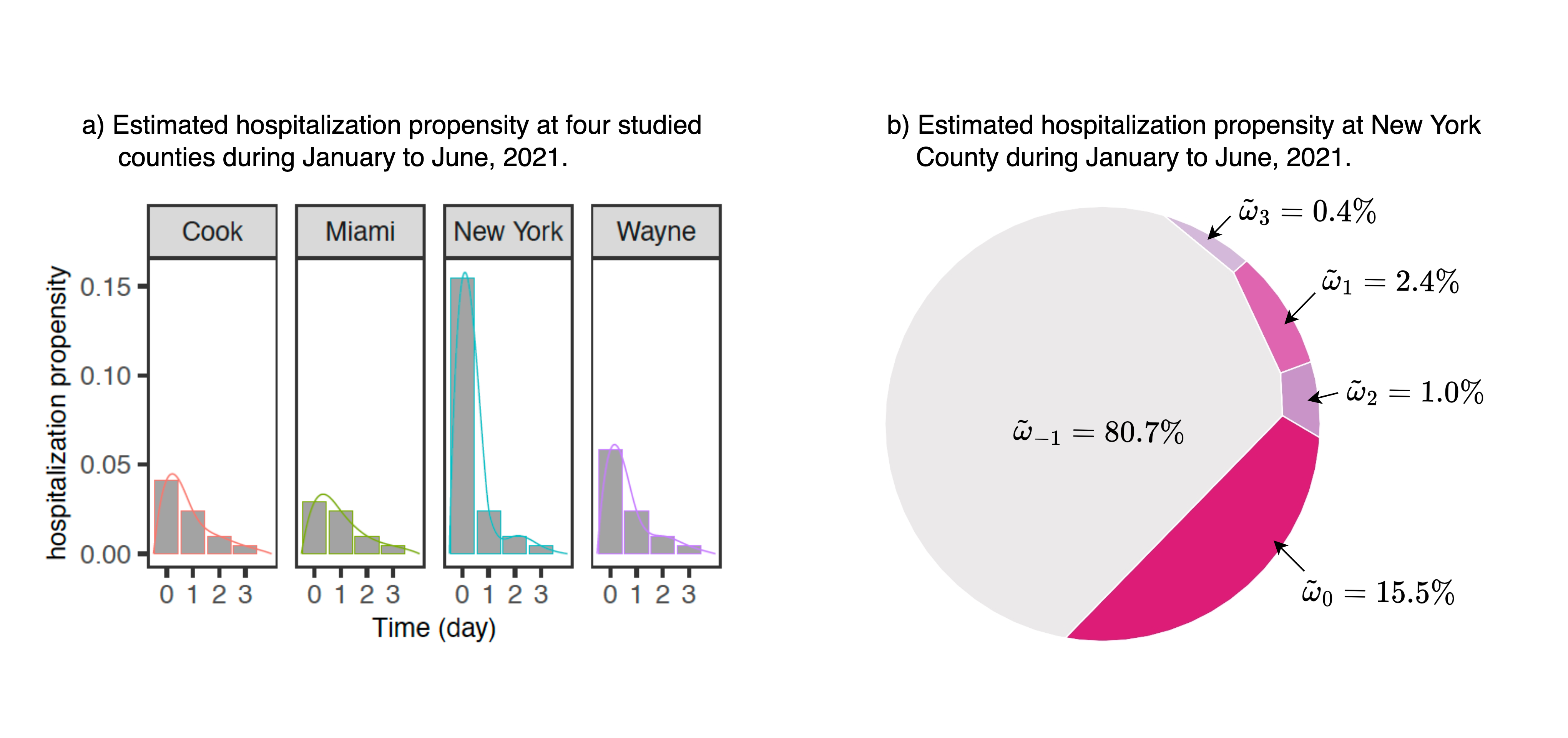}
		\caption{Visualization of the estimated hospitalization propensity, $\tilde\omega_s$, for COVID-19 in the four studied counties during February to May 2021. In panel a), the bar plot shows the estimated value of $\tilde\omega_s$ $1\leq s\leq 4$, at the four studied counties, while the solid lines represent the estimated propensity after fitting a cubic smoothing spline. In panel b), the estimated $\tilde\omega_s$ values at New York County are visualized using a Voronoi diagram to compare the magnitude of each $\tilde\omega_s$. Most of the hospitalized patients were admitted to a hospital on the day of infection.}
		\label{fig:Case5TOmega}
	\end{center}
\end{figure}

The final model selected was the same AR(1) structure used in the simulation study, with $\eta$ and $\tilde{\eta}$ estimated at 22 and 4, respectively. The estimated effect sizes were $0.1539$ (95\% CI: 0.1537 to 0.1541) for social distancing and $-7.3\times10^{-4}$ (95\% CI: $-8.3\times10^{-4}$ to $-6.4\times10^{-4})$) for temperature, respectively. These results suggested that lack of social distancing was a strong risk factor for elevated disease transmission during the study period. For example, a 50\% reduction in the frequency of visiting non-essential businesses was estimated to reduce $R_t$ by an average of 7.4\%. On the contrary, temperature exerted only a minor effect on disease transmission. The estimated county-level $R_t$ exhibited a similar trend during the study period among all four counties, as depicted in Figure S8. A decrease in disease transmission was observed since April 2021, which, in this dataset, was largely attributed to a reduction in social distancing value. Figure S9 illustrates the estimated infectiousness function. Its shape resembled the probability density function of either a Gamma or Weibull distribution (Figure S9a), with nearly two-thirds of secondary infections being diagnosed within the first week after the diagnosis of the infectors (Figure S9b). If we assume that the duration between infection and diagnosis is roughly the same for both infectors and secondary infections, our finding suggests that timely testing and a subsequent week-long quarantine of infected individuals can significantly mitigate disease transmission. The estimated hospitalization propensity for each county is presented in Figure \ref{fig:Case5TOmega}. Given the varied access to healthcare, hospitalization propensity diverged across locations. New York exhibited the highest propensity for hospitalization post-diagnosis (just under 20\%), compared to about 10\% in Miami. Hospitalization on the first day of diagnosis was also highest in New York (15.5\%) but below 5\% in Miami and Cook (Figure \ref{fig:Case5TOmega}a). Across counties, most hospitalizations occurred within 4 days of diagnosis, with an average of 1 day from diagnosis to admission. Assuming a 48-hour delay from symptom onset to diagnosis, most hospitalizations occurred within 6 days of symptom onset, with a mean time to admission of approximately 3 days. These findings align with prior studies. For example, a CDC report estimated that 20.9\% of U.S. COVID-19 patients were hospitalized before March 28, 2020 \citep{cdc2020preliminary}. \cite{zhang2020evolving} found a decrease in mean time from symptom onset to admission in Hubei, China, from 4.4 days (Dec 24–Jan 27, 2019) to 2.6 days (Jan 28–Feb 17, 2020). Traditionally, studying such durations requires epidemiological studies with contact tracing data, which poses a high requirement for the US disease surveillance system. Our results demonstrate that, even without extensive contact tracing, U.S. surveillance data can be effectively leveraged to estimate critical parameters for hospital planning and outbreak response.

\vspace{-10mm}
\section{Discussion}
\label{discussion}
%Given the existing importance of the TSI model in studying infectious diseases, the proposed method and algorithm elevate its usefulness and potential to a higher level. The proposed model not only enhances the robustness of the TSI model by incorporating hospitalization data but also enables investigations into infectiousness functions and hospitalization propensities. This, in turn, facilitates the prediction of hospitalization using the TSI model.

%In summary, we proposed an enhanced TSI model by incorporating hospitalization data and developed an estimation procedure using the composite likelihood approach and the MCEM algorithm. This refined model allows for more accurate estimation of the instantaneous reproduction number, as well as the estimation of the infectiousness function and hospitalization propensity without relying on contact tracing data. Additionally, it enables the assessment of the impact of risk factors. The proposed estimator improves the performance of TSI models, particularly when hospitalization data is of higher quality than incidence data, which is often the case in practice.

This study introduces a new model and estimation procedure that extends traditional TSI models by incorporating data augmentation for both incidence and hospitalization data, enabling more accurate estimation of the instantaneous reproduction number, especially when hospitalization data is more reliable than incidence data. Our model also facilitates estimating hospitalization propensity, previously achievable only through contact tracing, and assessing the association between risk factors and transmission dynamics. The model is broadly applicable where both incidence and hospitalization data are available, particularly when incidence data quality is low. %A key limitation is the computational time, which ranges from 1.7 to 6.5 minutes compared to 0.06 minutes for the approach in \cite{cori2013new}, though this delay is minimal given that $R_t$ updates are usually required daily or weekly.

The proposed method offers several extension opportunities. First, it can incorporate additional data, such as daily death counts, PCR tests, and serological data, to improve parameter accuracy and model death counts. Second, spatial correlations in disease transmission could be added using traffic data to enhance model efficiency. Third, the model can accommodate evolving pathogen dynamics and immunity changes, enabling analysis of new virus variants and immunity impacts. Fourth, it could be adjusted to handle overdispersion in infection counts by using distributions like the negative binomial, as illustrated in the supplementary materials. Finally, while the algorithm's runtime is manageable (1.7 to 6.5 minutes compared to 0.06 minutes in \cite{cori2013new}), further optimization could improve efficiency.

In addition to these extensions, the selection of appropriate risk factors is critical for applying our method effectively. In our study, we identified relevant factors through a systematic literature review and consultations with domain experts, focusing on those that significantly influence COVID-19 transmission dynamics. This approach ensured that the model captures local variations in transmission and improves the accuracy of $R_t$ estimation. Including irrelevant variables could introduce noise and bias, undermining the model's performance. To assist users in selecting relevant factors in practice, we propose the following practical guidelines: 1) conduct a comprehensive literature review and consult with domain experts to identify strong candidate risk factors; 2)ensure the availability of reliable data and assess the correlation between potential risk factors and $R_t$; 3)void multicollinearity and focus on variables that are most relevant to $R_t$.

\vspace{-4mm}
\section{Software}
\label{software}
Software in the form of R code, together with a sample example for the simulation study and complete documentation for the COVID-19 application is available on https://github.com/Jiasheng-Shi/Infectious-Disease-Hospitalization. Additional requests for implementation can be directed to the corresponding author at jing14@pennmedicine.upenn.edu.

\vspace{-4mm}
\section{Supplementary Material}
%\label{sec6}
Supplementary material is available online at
\url{http://biostatistics.oxfordjournals.org}.

\vspace{-4mm}
\section*{Acknowledgments}
Funding for the project was provided by the NIH
under award R01HD099348 and CDC under award U01CK000674. 
{\it Conflict of Interest}: None declared.

\bibliographystyle{biorefs}
\bibliography{ref}

\begin{table}[H]
	\def~{\hphantom{0}}
	\caption{Performance of the proposed method when the reported number of daily new infections was accurate, the infectiousness function $\omega_s$ and hospitalization propensity $\tilde\omega_s$ were unknown and no prior knowledge was available. For $\omega_s$ and $\tilde\omega_s$, only estimates of selected parameters are presented, due to the large number of parameters. Estimates of the other parameters not shown here yield similar results.} {%
		\begin{tabular}{lccccccc}
			\hline
			& $\theta_0$  & $\theta_1$ & $\beta_1$ & $\beta_2$ & $\omega_5$ & $\omega_6$ & $\tilde{\omega}_0$  \\[5pt]
			\hline
			%& U.S. population & White Rs & Black Rs & Hispanic Rs \\[2pt]
			Empirical bias ($\times 10^{-3}$)& -0.15 & 2.10 & 0.08 & 0.15 & -0.45 & -0.40 & -0.04 \\
			Relative bias ($\times 10^{-3}$) & -0.22 & 4.20 & -3.79 & -1.23 & -4.40 & -4.01 &-0.32 \\
			Standard error($\times 10^{-3}$) & 6.29 & 3.34 & 0.23 & 0.39 & 2.23 & 2.29 &0.60 \\
			95\% Coverage probability & 94.6\% & 92.0\% & 94.2\% & 96.2\% & 96.2\% & 97.0\% & 94.4\% \\
			\hline
	\end{tabular}}
	\label{estimationResults}
\end{table}

%\begin{figure}[!p]
%	\begin{center}
%		\includegraphics[width=1\textwidth]{figs/Fig3R2Q17.jpeg}
%		\caption{Estimation of the daily instantaneous reproduction number, $R_t$, when the reported number of daily new infections was accurate, the infectiousness function $\omega_s$ and hospitalization propensity $\tilde\omega_s$ were unknown and no prior knowledge was available. Black dotted line stands for the oracle $R_t$. Red solid line and the grey shadow stand for the estimates and its corresponding Bootstrapping confidence interval ($90\%$ confidence level) using the proposed method.}
%		\label{fig:EstR}
%	\end{center}
%\end{figure}

%\begin{figure}[!p]
%	\begin{center}
%		\includegraphics[width=1\textwidth]{figs/R_4counties}
%		\caption{Estimated instantaneous reproduction number, $R_t$, for COVID-19 in the four studied counties during February to May 2021. The blue lines represent the critical value of $R_t=1$. Red lines represent the estimated county-level $R_t$ during the study period. Shaded areas indicate the bootstrap confidence intervals ($90\%$ confidence level).
%		}
%		\label{county level: estimates}
%	\end{center}
%\end{figure}

%\begin{figure}[!p]
%	\begin{center}
%		\includegraphics[width=1\textwidth]{figs/Fig6_updated.png}
%		\caption{Visualization of the estimated infectiousness function, $\omega_s$, for COVID-19 in the four studied counties during February to May 2021. In panel a), the red dots represent the estimated $\omega_s$, $1\leq s\leq 22$, while the black line represents the estimated infectiousness function after fitting a cubic smoothing spline. In panel b), the estimated $\omega_s$ values are visualized using a Voronoi diagram to compare the magnitude of each $\omega_s$. Values of $\omega_s$ in the first week since infection are labeled.}
%		\label{fig:Case5Omega}
%	\end{center}
%\end{figure}

\end{document}

\def\spacingset#1{\renewcommand{\baselinestretch}%
{#1}\small\normalsize} \spacingset{1}

%%%%%%%%%%%%%%%%%%%%%%%%%%%%%%%%%%%%%%%%%%%%%%%%%%%%%%%%%%%%%%%%%%%%%%%%%%%%%%

\if1\blind
{
  \title{\bf  Time-Since-Infection Model for Hospitalization and Incidence Data}
  \author{Jiasheng Shi\thanks{
    The authors gratefully acknowledge the support provided by the Centers for Disease Control and Prevention (CDC) and the National Institutes of Health (NIH) under grants U01CK000674 and R01HD099348.}\hspace{.2cm}\\
    School of Data Science,\\ The Chinese University of Hong Kong- Shenzhen \vspace{2mm}\\Yizhao Zhou \\
    AstraZeneca\\
    and 
   \vspace{2mm}\\ Jing Huang \\
    Department of Biostatistics, Epidemiology and Informatics, \\Perelman School of Medicine at the University of Pennsylvania}
  \maketitle
} \fi

\if0\blind
{

  \bigskip
  \begin{center}
    {\LARGE\bf  Time-Since-Infection Model for Hospitalization and Incidence Data}
\end{center}
  \medskip
} \fi
\setcounter{figure}{7}  
\setcounter{algorithm}{1}  
\begin{center}
\section*{SUPPLEMENTARY MATERIALS}
\end{center}
\label{SM}
The Supplementary Materials include the following sections: (i) a general model setting where $D_{2,t}$ in ({\color{red}5}) is replaced with $\big\{ I_r,  H_r \big\}_{0\leq r\leq t-1}$, along with its corresponding simulation performance; (ii) an extension of the model to account for overdispersion in daily infection counts; (iii) Algorithm 2, an MCEM algorithm for parameter estimation when prior knowledge of $\omega$ and $\omega_s$ is available; (iv) additional clarification on the simulation setup; (v) a comparison of the estimated $R_t$ with that of \cite{cori2013new} using various sliding window lengths; (vi) simulations addressing underreported incidence and hospitalization data; (vii) detailed proofs of all technical results; and (viii) additional results from the simulation study and application to COVID-19 Data.
\vskip 1em 

\subsection{General model setting and corresponding simulation}
%\subsection{General setting of the latent time series of the instantaneous reproduction number}

Consider the general setting in which  we replace $D_{2,t}$ in ({\color{red}4}) and ({\color{red}5}) with $\{ I_j, H_j   \}_{0\leq j\leq t-1}$, denoted as $D_{2,t}'$. Since $D_{2,t}' \subset \mathcal{G}_{t}$, we will first make adjustment to the base layer of the TSI model in Equation ({\color{red}1}) to 
\begin{equation}
I_t \big| \mathcal{G}_{t-1}  \; \sim \; {\rm Poisson}( R_t \Lambda_{t}), \quad {\rm with}\;\; \Lambda_t=\sum_{1\leq s\leq t} \omega_s I_{t-s}. \label{countdistributionVersion2} \tag{12}
\end{equation}
In most cases (with the general formula in Equation ({\color{red}4})), the joint likelihood and composite likelihood become intractable due to the irregular range restriction of $H_r \mid \mathcal{G}_{r-1}$. However, for a special case of Equation ({\color{red}4}) where $R_t$ depends on $D_{2,t}'$ only through $\{ H_{t-1} \}\cup \mathcal{F}_{t}$, and $R_t$ is independent of $\{ H_{t-2},\cdots, H_0 \}$, e.g.,
\begin{equation}
\log(R_{t}) - \theta_1\log \big( H_{t-1}/I_{t-1}   \big)=Z_{t}^T\beta  + \theta_2 \Big[ \log \big(R_{t-1} \big) -\theta_1\log \big( H_{t-2}/I_{t-2}   \big) \Big] \label{GTS} \tag{13}.
\end{equation}
In practice, the term involving $H_{t-1}/I_{t-1}$ can be interpreted as the effect brought by quarantine or similar interventions, and we obtain  the composite likelihood
\begin{align*}
\ell_{C}&= \sum_{0\leq r\leq t } \log \mathbb{P}( H_r, I_r \mid \mathcal{F}_{r-1}) \\
&= \sum_{0\leq r\leq t } \log \mathbb{P}( H_r\mid \mathcal{F}_{r}) \cdot \mathbb{P}( I_r \mid \mathcal{F}_{r-1}) \\
&= \sum_{0\leq r\leq t } \log \mathbb{P}( H_r\mid \mathcal{F}_{r}) +  \sum_{0\leq r\leq t } \log  \sum_{\ddot{H}_{r-1}} \mathbb{P}(  I_r \mid \ddot{H}_{r-1},\mathcal{F}_{r-1} ) \cdot \mathbb{P}( \ddot{H}_{r-1} \mid \mathcal{F}_{r-1} ).
\end{align*}
Notice that $\mathbb{P}(  I_r \mid \ddot{H}_{r-1},\mathcal{F}_{r-1} )$ is a Poisson distribution function. For each $0\leq r\leq t$, 
\begin{align*}
\mathbb{P}(  \ddot{H}_r \mid \mathcal{F}_r ) &= \sum_{ h_{r-\tilde{\eta},\tilde{ \eta}}, \cdots, h_{r-1,1} } \mathbb{P} \Big(  h_{r,0}=\big(\ddot{H}_r - \sum_{s=1}^{\tilde{ \eta}} h_{r-s,s} \big) , h_{r-s,s}, 1\leq s\leq \tilde{ \eta} \mid \mathcal{F}_{r}    \Big) \\
&= \sum_{ h_{r-\tilde{\eta},\tilde{ \eta}}, \cdots, h_{r-1,1} }  \mathbb{P}\Big(Binomial(I_{r}, \tilde{ \omega }_0 ) =  \big(\ddot{H}_r - \sum_{s=1}^{\tilde{ \eta}} h_{r-s,s} \big) \Big)  \\
&\qquad \qquad \qquad \qquad \qquad   \cdot  \prod_{s=1}^{\tilde{ \eta}} \mathbb{P}\big(Binomial(I_{r-s}, \tilde{ \omega }_s ) = h_{r-s,s}  \big). 
%& = \e \Big[ \P\Big(Binomial(I_{r}, \tilde{ \omega }_0 ) =  \big(\ddot{H}_r - \sum_{s=1}^{\tilde{ \eta}} h_{r-s,s} \big) \Big) \mid \mathcal{F}_r \Big]
\end{align*}

We conduct a simulation using the general setting model, where we set the infectiousness function and the hospitalization propensity as
\begin{gather*}
\omega=(0.27,0.29,0.17,0.12,0.09,0.06), \;\; {\rm with} \;\; \eta=6, \;\; \\
{\rm and}\;\; \tilde{\omega}=(\tilde{\omega}_{-1}=0.85, \tilde{ \omega }_0=0.1, \tilde{ \omega }_1=0.05),  \;\; {\rm with} \;\; \tilde{\eta}=2. \;\; 
\end{gather*}
We further set the number of infections, hospitalizations, and the instantaneous reproduction number at time zero to be $50$, $7$ and $2.5$, respectively. The simulated study period lasts for $T=120$ days, and the maximum daily incidence is below $500$. We incorporate two covariates, social distancing and wet-bulb temperature, directly from the New York County data used in Section 4 of the main manuscript, covering the period from January 1st to April 30th, 2021, with covariate effects being $(0.25,-0.01)$. We estimate the model parameters and the instantaneous reproduction number over time and compare them to the true values. The results are similar to those shown in the simulation section of the main manuscript. We also demonstrate the estimated instantaneous reproduction number from one randomly selected replicate in Figure \ref{fig:Case6}. 

\renewcommand{\thefigure}{S1}
\begin{figure}[htbp!]
    \centering
	\includegraphics[width=1\textwidth]{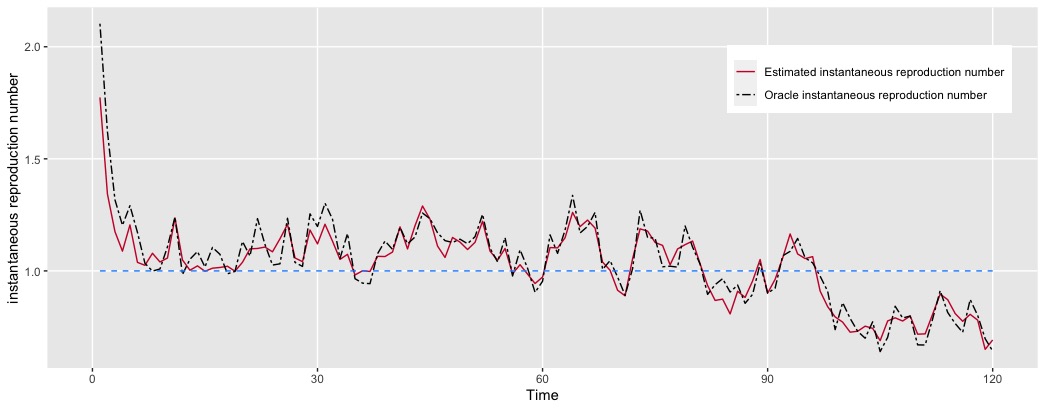}
	\caption{The estimated instantaneous reproduction number from one randomly selected replicate using data simulated from the general setting model.}
	\label{fig:Case6}
\end{figure}

\vskip 1em

\subsection{Model extention: overdispersion in daily infection numbers}
The proposed approach is applicable to distributions with an overdispersion parameter. However, the composite likelihood functions need to be adjusted on a case-by-case basis. For example, when $I_t \vert \mathcal{F}_{t-1}$ is assumed to follow a negative binomial distribution, i.e.,
\begin{gather*}
	%I_t \vert \mathcal{F}_{t-1} \sim {\rm Negative-Binomial}\left( r=R_t\Lambda_t, p  \right), \\
	\P\big(  I_t \vert \mathcal{F}_{t-1}  \big) =  \frac{\Gamma(I_t+ \Lambda_t R_t)}{ \Gamma( R_t \Lambda_t ) I_t!  } (1-p)^{I_t} p^{R_t \Lambda_t},
\end{gather*}	

the composite likelihood (2.8) in our manuscript needs to be updated accordingly. Conditional on $\mathcal{F}_{t-1}\triangleq \sigma\{ I_s, s\leq t-1 \}$,
\begin{align*}
	&\P( H_t, I_t , h_{t-\eta,\eta},\cdots, h_{t-1,1} \vert \mathcal{F}_{t-1} ) \\
	=& \P \bigg( h_{t-\eta,\eta},\cdots, h_{t-1,1}, h_{t,0}=H_t-\sum_{s=1}^{\eta}h_{t-s,s}, \\
	&\qquad \qquad \qquad \qquad  \sum_{s=-1,1,\cdots,\eta} h_{t,s}=I_t-H_t+ \sum_{s=1}^{\eta}h_{t-s,s} \vert \mathcal{F}_{t-1}  \bigg)  \\
	=& \prod_{s=1}^{\eta} \P \Big( Binomial(I_{t-s},\tilde{\omega}_s) \Big) \cdot \mathbbm{1} \Big( H_t-I_t \leq \sum_{s=1}^{\eta} h_{t-s,s} \leq H_t   \Big) \\
	& \qquad  \times  \P \bigg( Multinomial\bigg( I_t+R_t\Lambda_t \; ;\;  \tilde{\omega}_{0}(1-p) ,  (1-\tilde{\omega}_{0})(1-p) , p \bigg) \\
	& \qquad  = \big( h_{t,0}=H_t-\sum_{s=1}^{\eta}h_{t-s,s}, \sum_{s=-1,1,\cdots,\eta} h_{t,s}=I_t-H_t+ \sum_{s=1}^{\eta}h_{t-s,s}, R_t\Lambda_t  \big) \bigg).
\end{align*}
Finally,
\begin{equation*}
	\P(H_t,I_t\vert \mathcal{F}_{t-1}) = \sum_{h_{t-\eta,\eta},\cdots,h_{t-1,1}} \P \big( H_t,I_t, h_{t-\eta,\eta},\cdots,h_{t-1,1} \vert \mathcal{F}_{t-1} \big) .
\end{equation*}
Therefore, the MCEM approach and the proposed estimation method can still be applied, though the objective functions must be updated accordingly. 

\vskip 0.5em

\subsection{An MCEM algorithm to estimate the model parameters when prior knowlege about $\omega$ and $\tilde\omega$ is available.}

In practice, estimates of $\omega_s$ and $\tilde{\omega}_s$ might be available from small biomedical or contact tracing studies. Due to the diversity in sample sizes, the demographics of study cohorts, and the estimation methods employed in different studies, these existing estimates can vary widely. In such cases, it's possible to treat these estimates as prior knowledge and use them to estimate the model parameters with Algorithm~\ref{MCEM2}. In this algorithm, we assume that $\omega_s$ and $\tilde{\omega}_s$ can be approximated using the probability density function of gamma distributions, such that
\begin{gather*}
	\omega_s(c)= \mathbb{P} \Big( s-1\leq  \text{Gamma}(k_1(c),\mu_1(c)) \leq s    \Big), \quad  s=1,\cdots, \eta-1,\\
	%{\rm and}\quad \omega_{\eta}(c) = 1- \sum_{s=1}^{\eta-1} \omega_s (c), \\
	\tilde{\omega}_s(c)= \mathbb{P} \Big( s\leq  \text{Gamma}(k_2(c),\mu_2(c)) \leq s+1    \Big), \quad  s=0,\cdots, \tilde{ \eta }, \\
	\text{and}\quad \omega_{\eta}(c) = 1- \sum_{s=1}^{\eta-1} \omega_s (c), \quad  \tilde{\omega}_{-1}(c) = 1- \sum_{s=1}^{ \tilde{\eta} } \tilde{\omega}_s(c),
\end{gather*}

where  $(k_1(c),\mu_1(c),k_2(c),\mu_2(c))$ are the shape and rate parameters of the gamma distribution and $c\in \mathcal{C}$ is the index of prior studies that generated the estimates of $\omega_s$ and $\tilde\omega_s$. We propose to assign probability weights $W_c$ to each candidate estimate of  $\omega$ and $\tilde\omega$ and assume that the true values of $\omega_s$ and $\tilde\omega_s$ can be calculated as a weighted average from these studies. A pseudo-code algorithm is shown in Algorithm \ref{MCEM2}.

\begin{algorithm}
	\caption{An MCEM algorithm for estimating the model parameters when prior knowlege of $\omega_s$ and $\tilde\omega_s$ is  available.}
	\label{MCEM2}
	\begin{algorithmic}
		\State Assign probability weights $W_c$, $c\in \mathcal{C}$, to candidate estimate $(\omega(c),\tilde{ \omega }(c))$.
		\For {$c\in\mathcal{C}$, }
		\Require initial parameter value $\gamma^{(0)}(c)=(\beta^{(0)}(c),\theta^{(0)}(c),\omega(c),\tilde{ \omega }(c))$, $k\leftarrow0$,
		and the breakpoint critical value $\Delta_0$.
		\While {$ \Vert \gamma^{(k+1)}(c) -\gamma^{(k)}(c) \Vert_{\infty} > \Delta_0  $, }
		\For {$0\leq r\leq t$, $1\leq m \leq N_0$,} 
		\State Sample $Data_{miss,r}^{(m,k)}$ from $\mathbb{P}_{\gamma^{(k)}(c)} \big(H_r, I_r, Data_{miss,r} \mid \mathcal{F}_{r-1} \big) $ derived in ({\color{red}8})
		\EndFor
		\State Calculate the Monte Carlo Q-function $\hat{Q}(\gamma \mid \gamma^{(k)}(c))$,
		\State Seek for the estimate $\gamma^{(k+1)}(c) = \arg\max_{\gamma \in \Gamma} \hat{Q}(\gamma\mid \gamma^{(k)}(c))\vert_{\omega=\omega(c),\tilde{ \omega }=\tilde{ \omega }(c)}.$
		\State let $k\leftarrow k+1$.
		\EndWhile
		\State Take the estimate from the final iteration $\gamma^{(k+1)}(c) $ as $\gamma(c)$.
		\EndFor
		\State Output the final estimate, which is given by $\hat{\gamma} = (\sum_{c\in\mathcal{C}} W_c\gamma(c)) / ( \sum_{c\in\mathcal{C}} W_c )$.
	\end{algorithmic}
\end{algorithm}

In our local data generating process, we used an infectiousness function that mimicked the reported serial interval from \cite{li2020early}. However, for the data analysis, we applied an infectiousness function derived from a meta-analysis of results from \cite{wu2020estimating,ali2020serial,chen2022inferring,deng2021estimation}, using it as the initial value for estimating the infectiousness function parameters. We presenting the information here for reference,
\vskip 0.5em
For infectiouness function:
\begin{enumerate}
	
	\item In \cite{li2020early}, the estimated infectiousness function follows a Gamma distribution with a mean of 7.5 days (95\% CI, [5.3, 19]). This corresponds to a standard deviation ranging approximately from 1.12 to 5.86. In our simulation, we set the infectiousness profile using a Gamma distribution with shape and scale parameters of 2.5 and 3, respectively, resulting in a mean of 7.5 and a standard deviation of 4.74.
	%\item The pooled infectiousness function based on Wu et al., 2020; Ali et al., 2020; Chen et al., 2022; and Deng et al., 2021, {\color{red}is xxx}. The estimated infectiousness function from each individual study is as follows.
	\item In \cite{wu2020estimating}, the estimated infectiousness function has a gamma distribution with mean $7.0$ and standard deviation $4.5$.
	%	\begin{gather*}
		%		k= (7/4.5)^2 \approx 2.42, \quad \theta = 7/k \approx 2.89, \\
		%		 \omega_s = \P\big( \Gamma(2.42, 2.89) \leq s  \big) - \P\big( \Gamma(2.42, 2.89) \leq s-1  \big).
		%	\end{gather*}
	\item In \cite{ali2020serial}, the estimated serial interval (that we used to approximate the infectiousness function) has a normal distribution with mean $5.1$ and standard deviation $5.3$. %So we used the truncated normal distribution for infectiousness profile with the same mean and variance.
	\item In \cite{chen2022inferring}, the estimated temporal generation time distribution (that we used to approximate the infectiousness function) is given by Weibull distribution with mean $5.20$ and standard deviation $3.65$.
	\item In \cite{deng2021estimation}, the estimated incubation period (that we used to approximate the infectiousness function)  is given by a gamma distribution with mean 2.73 and standard deviation 1.23. %$(\alpha,\beta)=(4.97,0.55)$.
\end{enumerate}
Similarly, our hospitalization propensity was set using a Gamma distribution to mimic the distribution of the delay between illness onset and hospitalization reported in \cite{li2020early}.
\vskip 1em

\subsection{Additional clarification on the simulation process and guidance on model selection for real-world applications}

The simulated COVID-19 incidence curve and social distancing variable, designed to mimic the Unacast data, are shown below in Figure \ref{fig:visualisation}.

\renewcommand{\thefigure}{S2}
\begin{figure}[htbp!]
\centering	
\includegraphics[width=0.9\textwidth]{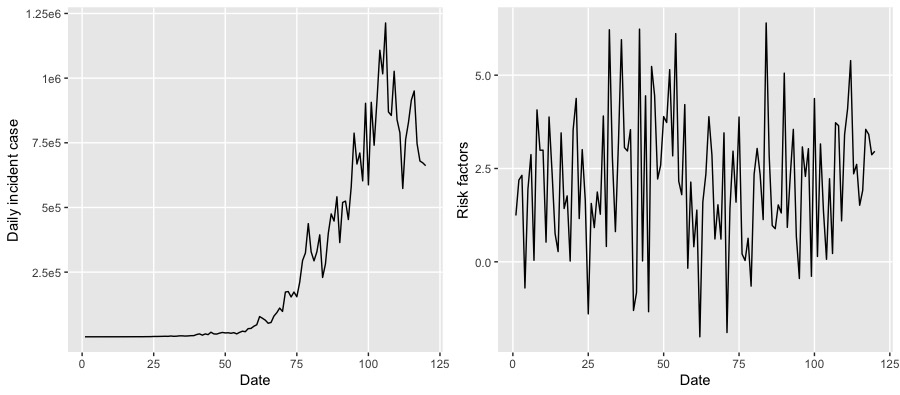}
\caption{The left panel shows the simulated COVID-19 incidence curves, while the right panel displays simulated risk factor data, representing the percentage change in visits to nonessential businesses (based on Unacast data) after applying a logistic transformation.}
\label{fig:visualisation}
\end{figure}

In our real data application, the model was selected using the following procedure:

{\underline{Selection of risk factors}:
The risk factors in our model were selected based on a thorough review of the literature that has extensively examined local factors influencing COVID-19 transmission. Studies such as \cite{talic2021effectiveness} on the effectiveness of social distancing measures, \cite{weaver2022environmental} on environmental determinants, and \cite{rubin2020association} on the role of population density, among others, have consistently shown that factors like social distancing, population density, and environmental conditions (such as temperature) are significantly associated with disease transmission. Although temperature may have a smaller effect, its influence is still relevant and was therefore included in our analysis.

When using the proposed method, including variables that are genuinely associated with $R_t$ enhances model performance, while including unrelated variables can introduce noise and reduce efficiency, as they do not improve estimation accuracy nor aid in identifying relevant risk factors. While risk factor selection is not the primary theoretical focus of the method, it is an important practical component in real-world applications. To assist users in risk factor selection, we propose the following practical guidelines:
\begin{itemize}
	\item Conduct a comprehensive literature review and consult with domain experts to identify strong candidate risk factors.
	\item Ensure the availability of reliable data and assess the correlation between potential risk factors and $R_t$.
	\item Avoid multicollinearity and focus on variables that are most relevant to $R_t$.
\end{itemize}
\vskip 1em

{\underline{Model selection}:
	We focus on the discussion of model diagnostics and model selection for the time series component of the proposed method. In our data application, we applied the Box-Cox transformation to the preliminary $R_t$ estimates obtained using Cori's method to determine an appropriate link function. The estimated lambda values for each county ranged from -1 to 0.5, with an overall estimate between 0 and 0.5, leading us to use the log link function. Subsequently, we fitted a generalized linear mixed-effects model with the $R_t$ estimated by Cori's method as the outcome and the selected risk factors as covariates using a log link function. After applying the log link function, we calculated the residuals and examined the partial autocorrelation function (PACF) of the residuals by county to determine the appropriate order for an autoregressive (AR) model. The PACF plot indicated an AR(1) structure for three of the four counties and an AR(2) structure for one county. For model parsimony, we opted to use an AR(1) structure for the overall model.
	
In general, one may begin by obtaining preliminary estimates of the reproduction number from observed incidence data using Cori's estimator. A time series model can then be fitted to these estimates along with exogenous variables, following standard time series modeling procedures.
	\begin{itemize}
		\item  Data Transformation: Apply the Box-Cox transformation to stabilize variance over time, typically using a logarithmic transformation, as it often improves model fit in time series data.
		
		\item Decomposition: Decompose the time series into trend, seasonal, and noise components using STL (seasonal-trend decomposition via LOESS). The trend component can be estimated through non-parametric regression or a simple linear model. This decomposition allows for isolating the underlying structure from random fluctuations.
		
		\item Noise Modeling and ARIMA Selection: To model the noise component, use ACF and PACF plots to select the ARIMA model order, then refine the selection using criteria like AIC or BIC to balance fit and complexity.
		
		\item High-Dimensional Variable Selection: When working with many exogenous variables, an information-based selection method can help identify a relevant subset, minimizing model complexity while retaining the most significant covariates.
		
	\end{itemize}

\vskip 1em

\subsection{A comparison of the estimated $R_t$ to reference approach with different sliding window}

The Figure 2 uses a 3-day sliding window chosen from the criteria of minimizing the $L^2$ distance between the estimated and oracle sequences of $R_t$. We observed that $R_t$ estimated at the window midpoint generally had a smaller bias compared to $R_t$ estimated at the window endpoint, which was consistent with the results of \cite{gostic2020practical} and \cite{Gressani2023approximate}. However, for the majority of time points (e.g., 45 out of 57 days as shown in Fig.2, 78.9\%), our proposed method consistently produced estimates with smaller bias than Cori's method, regardless of whether $R_t$ was reported at the window's endpoint or midpoint. The magnitude of bias reduction is substantial, as shown in Figure 2.

\vskip 1em
{\underline{1-day sliding window}:
When using a 1-day sliding window, Cori's $R_t$ estimates performed similarly to or even slightly better than the proposed estimates in the later stage of the simulated period, where the daily incident cases $I_t$ were large enough ($\geq 5\times 10^4$) for Cori's method to estimate $R_t$ accurately without incorporating hospitalization data, as shown in Figure \ref{fig:estimationbiascomparison3}.  For both methods, the bias was of the order $O(10^{-2})$ during this period. 
	
	However, Cori's estimates showed a substantial bias in the early stage of the simulated period when the incident cases $I_t$ were small ($\leq 2\times 10^3$), while our proposed estimates maintained a small bias by incorporating hospitalization data, as shown in Figure \ref{fig:estimationbiascomparison2}. During this period, the proposed estimator showed consistent performance with bias of the order $O(10^{-2})$, whereas Cori's estimator had a much larger bias, on the order of $O(10^{-1})$ (or even $O(1)$). This suboptimal performance of Cori's method in the early stage is also why the data-driven window length selection resulted in a 3-day sliding window instead of a 1-day window. 
	
	In general, the 1-day sliding window limits Cori's method's ability to borrow information from nearby days, leading to unsatisfactory performance when daily incident cases $I_t$ are small. This drawback becomes even more pronounced when estimating $R_t$ in smaller populations, such as a county with roughly 10 cases per day. In contrast, our proposed method is less affected in these cases as it leverages both incidence and hospitalization data. 
\vskip 0.5em

\renewcommand{\thefigure}{S3}
\begin{figure}[ht!]
	\centering	
	\includegraphics[width=0.7\textwidth]{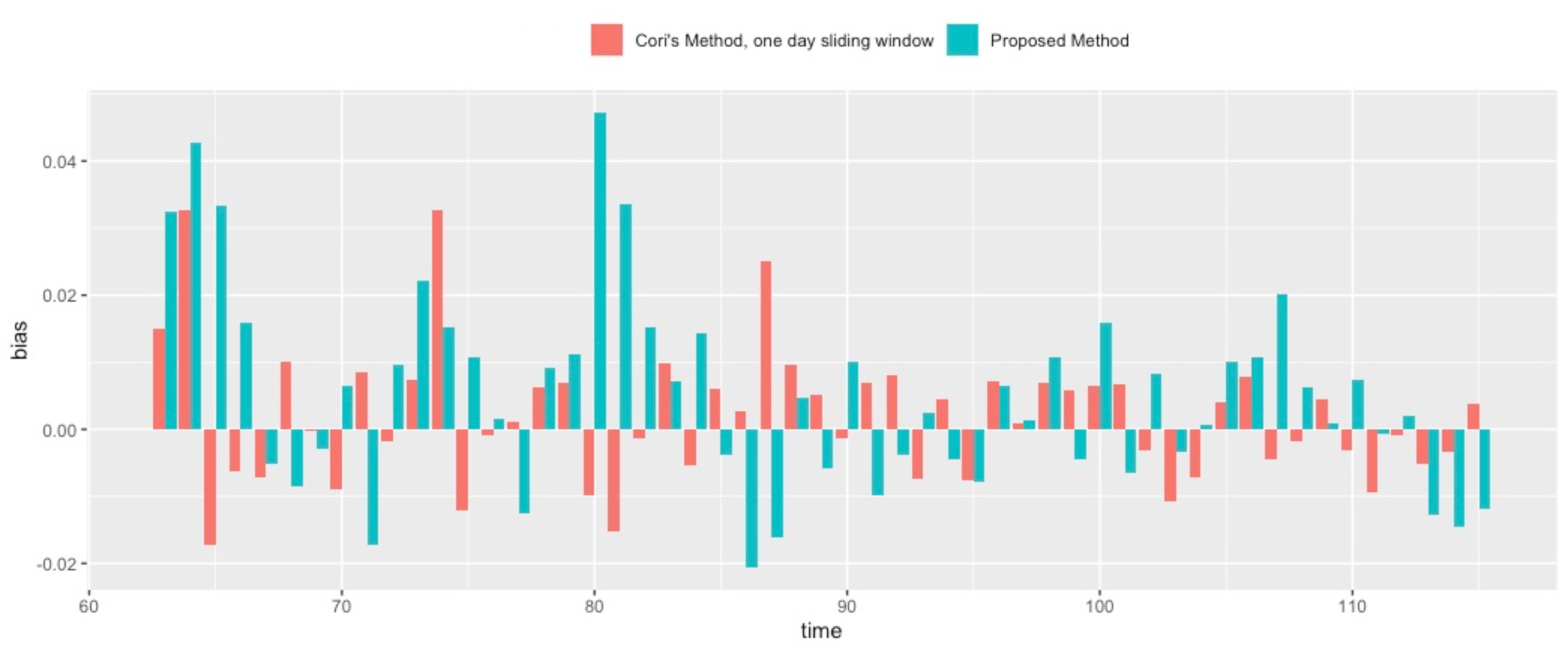}
	\vspace{-4mm}
	\caption{Comparison of the estimation bias of $R_t$ between the proposed method and Cori's method using a  1-day sliding window for Cori's method, during the later stage of the simulated period when $I_t$ is large. }
	\label{fig:estimationbiascomparison3}
\end{figure}

\renewcommand{\thefigure}{S4}
\begin{figure}[ht!]
	\centering	
	\includegraphics[width=0.66\textwidth]{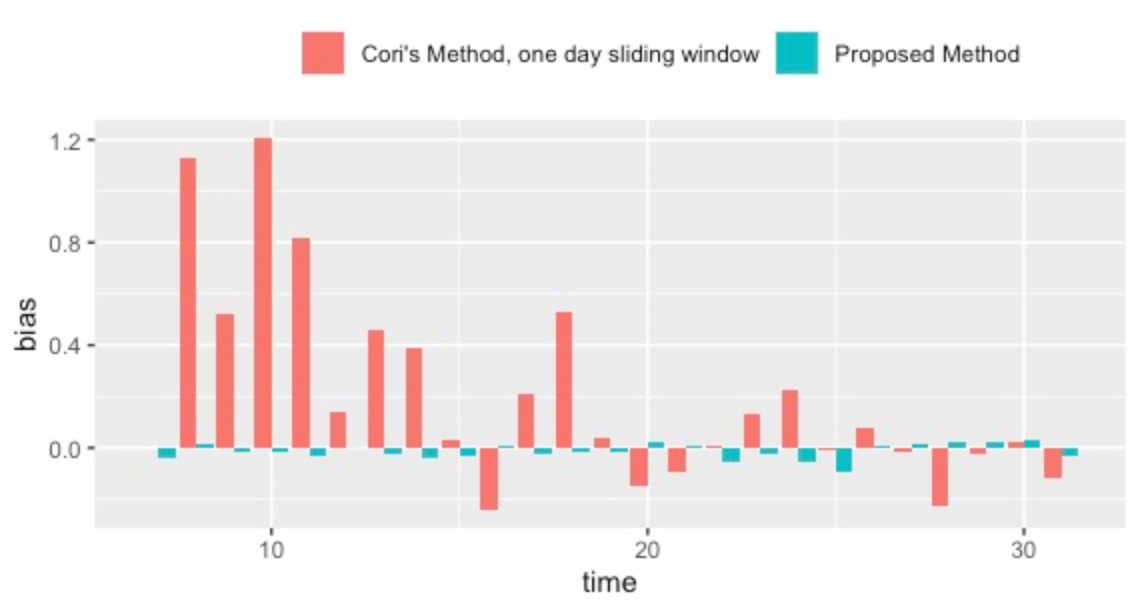}
	\vspace{-4mm}
	\caption{Comparison of the estimation bias of $R_t$ between the proposed method and Cori's method using a  1-day sliding window for Cori's method, during the early stage of the simulated period when $I_t$ is small.}
	\label{fig:estimationbiascomparison2}
\end{figure}

{\underline{7-day sliding window}: When using a 7-day sliding window, reporting $R_t$ at the midpoint for Cori's method resulted in smaller bias compared to reporting at the endpoint. However, the overall bias was much larger compared to using Cori's method with a 3-day sliding window or our proposed method, as the 7-day window tended to oversmooth the estimates. Given the significantly larger bias relative to our proposed method, we illustrate the comparison using a randomly selected simulated dataset, rather than a bar plot of the bias, as shown in Figure \ref{fig:estimationbiascomparison4}.

\renewcommand{\thefigure}{S5}
\begin{figure}[htbp!]
	\centering	
	\includegraphics[width=0.95\textwidth]{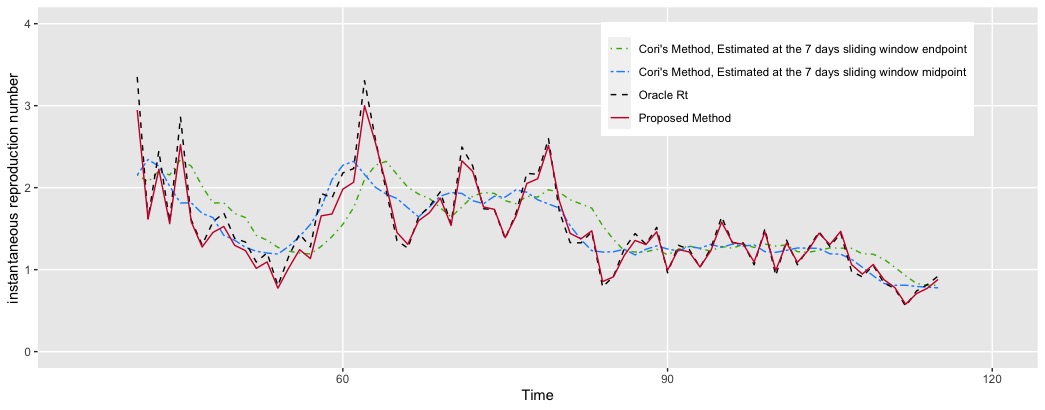}
	\vspace{-4mm}
	\caption{Comparison of the estimation bias of $R_t$ between the proposed method and Cori's method using a  7-day sliding window for Cori's method.}
	\label{fig:estimationbiascomparison4}
\end{figure}

\vskip 1em
\subsection{A simulation scenario with under-reported incidence and hospitalization data}

In this scenario, we generate the data where both incidence and hospitalization data were subject to under-reporting, with daily reporting rates fluctuating randomly between 0\% and 30\%. Under this condition, the estimation bias of $R_t$ increased in our proposed method, but remained significantly smaller than the bias observed in Cori’s method, as shown in Figure \ref{fig:Barchart}. This indicates that the proposed method retains a degree of robustness to data inaccuracies, even when the hospitalization data are imperfect.

\renewcommand{\thefigure}{S6}
\begin{figure}[htbp!]
	\centering	
	\includegraphics[width=0.9\textwidth]{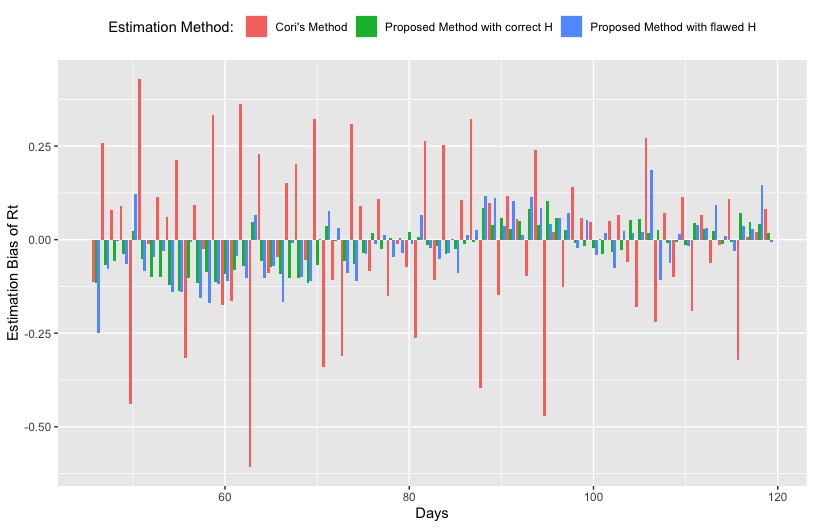}
	\vspace{-5mm}
	\caption{Comparison of estimation bias in $R_t$ using under-reported incidence and hospitalization data.}
	\label{fig:Barchart}
\end{figure}

\subsection{Detailed proof of technique results}
\label{proof}

\begin{proof}{of Lemma {\color{red}1}.}
	According to ({\color{red}1}) and ({\color{red}3}), notice that, conditional on $ \mathcal{F}_{t-1} \triangleq \sigma \{ I_{s}, s\leq t-1 \}$, 
	\begin{align*}
	&\mathbb{P} ( h_{t,u_1}, h_{t,u_2} | \mathcal{F}_{t-1}  ) = \sum_{I_t} \mathbb{P} ( h_{t,u_1}, h_{t,u_2} | I_t ) \cdot \mathbb{P}( I_t |  I_0,\cdots, I_{t-1} )  \label{cond-dist-poi} \tag{14}\\
	=& \sum_{I_t=h_{t,u_1}+h_{t,u_2}}^{\infty} \bigg[  \frac{I_t ! }{h_{t,u_1}! h_{t,u_2}! } \tilde{ \omega }_{u_1}^{h_{t,u_1}} \tilde{ \omega }_{u_2}^{ h_{t,u_2} } \cdot \frac{ (1-\tilde{ \omega }_{u_1}-\tilde{ \omega }_{u_2})^{I_t-h_{t,u_1}-h_{t,u_2}} }{  \big(  I_t-h_{t,u_1}-h_{t,u_2} \big)!  }   \\
	&\qquad \qquad \qquad \qquad  \cdot   \frac{(R_{t}\Lambda_t)^{I_t}}{I_t !} e^{-R_{t}\Lambda_t} \bigg] \\
	=&  \frac{ (\tilde{ \omega }_{u_1} R_{t}\Lambda_t )^{h_{t,u_1}} }{h_{t,u_1}!} e^{-\tilde{ \omega }_{u_1} R_{t}\Lambda_t } \cdot  \frac{ (\tilde{ \omega }_{u_2} R_{t} \Lambda_t )^{h_{t,u_2}} }{h_{t,u_2}!} e^{-\tilde{ \omega }_{u_2} R_{t}\Lambda_t } \\
	& \qquad \cdot \sum_{I_t=h_{t,u_1}+h_{t,u_2}}^{\infty}   \frac{ \big(  (1- \tilde{ \omega }_{u_1}- \tilde{ \omega }_{u_2}  )  R_{t} \Lambda_t \big)^{ I_t-h_{t,u_1}-h_{t,u_2} } }{  \big(   I_t-h_{t,u_1}-h_{t,u_2}   \big) !} e^{-(1- \tilde{ \omega }_{u_1}- \tilde{ \omega }_{u_2}  )  R_{t}\Lambda_t } \\
	=&  \frac{ (\tilde{ \omega }_{u_1} R_{t}\Lambda_t )^{h_{t,u_1}} }{h_{t,u_1}!} e^{-\tilde{ \omega }_{u_1} R_{t}\Lambda_t } \cdot  \frac{ (\tilde{ \omega }_{u_2} R_{t} \Lambda_t )^{h_{t,u_2}} }{h_{t,u_2}!} e^{-\tilde{ \omega }_{u_2} R_{t}\Lambda_t }. 
	\end{align*}
	Hence ({\color{red}7}) holds and $h_{t,u_1}|\mathcal{F}_{t-1} \perp h_{t,u_2} |\mathcal{F}_{t-1}$. Besides, it's clear that $(h_{t,u_1}, h_{t,u_2} ) | \mathcal{G}_{t-1}=(h_{t,u_1}, h_{t,u_2} ) | \mathcal{F}_{t-1}$ from Figure.{\color{red}1}, thus proves Lemma {\color{red}1}.
\end{proof}

\vspace{1em}

\begin{proof}{of Theorem {\color{red}1}.}
	Notice that 
	\begin{equation*}
	\log \mathbb{P}_{\gamma} \big(H_r, I_r |\mathcal{F}_{r-1} \big) = \log \mathbb{P}_{\gamma} \big(H_r, I_r, Data_{miss,r} |\mathcal{F}_{r-1} \big) - \log  \mathbb{P}_{\gamma} \big( Data_{miss,r} |\mathcal{D}_{obs,r} \big) 
	\end{equation*}
	therefore,
	\begin{align*}
	\log \mathbb{P}_{\gamma} \big(H_r, I_r |\mathcal{F}_{r-1} \big) &= \int \log \mathbb{P}_{\gamma} \big(H_r, I_r |\mathcal{F}_{r-1} \big) d \mathbb{P}_{\gamma^{(k)}}  \big( Data_{miss,r} |\mathcal{D}_{obs,r} \big) \\
	&= \int \log \mathbb{P}_{\gamma} \big(H_r, I_r, Data_{miss,r} |\mathcal{F}_{r-1} \big) d \mathbb{P}_{\gamma^{(k)}}  \big( Data_{miss,r} |\mathcal{D}_{obs,r} \big) \\
	& \qquad - \int  \log  \mathbb{P}_{\gamma} \big( Data_{miss,r} |\mathcal{D}_{obs,r} \big) d \mathbb{P}_{\gamma^{(k)}}  \big( Data_{miss,r} |\mathcal{D}_{obs,r} \big) \\
	&\triangleq \mathbb{E}_{\gamma^{(k)}} \Big(   \log \mathbb{P}_{\gamma} \big(H_r, I_r, Data_{miss,r} |\mathcal{F}_{r-1} \big) \vert \mathcal{D}_{obs,r}   \Big) - H_r( \gamma \vert \gamma^{(k)} )
	\end{align*}
	Since $H_r (  \gamma \vert \gamma^{(k)} ) \leq H_r ( \gamma^{(k)} \vert \gamma^{(k)})$ for arbitrary $\gamma\in \Gamma$ and $0\leq r\leq t$ according to the information inequality (Gibbs inequality), hence
	\begin{align*}
	\ell_C (\gamma^{(k+1)}) - \ell_C (\gamma^{(k)}) &= \sum_{0\leq r\leq t} \log \mathbb{P}_{\gamma^{(k+1)}} \big(H_r, I_r |\mathcal{F}_{r-1} \big) -  \sum_{0\leq r\leq t} \log \mathbb{P}_{\gamma^{(k)}} \big(H_r, I_r |\mathcal{F}_{r-1} \big)   \\
	&= Q(\gamma^{(k+1)} \vert \gamma^{(k)}) - Q(\gamma^{(k)} \vert \gamma^{(k)}) +    \sum_{0\leq r\leq t}  \Big(  H_r (\gamma^{(k)} \vert \gamma^{(k)}) -  H_r (\gamma^{(k+1)} \vert \gamma^{(k)})  \Big)  \\
	& \geq Q(\gamma^{(k+1)} \vert \gamma^{(k)}) - Q(\gamma^{(k)} \vert \gamma^{(k)})  \geq 0,
	\end{align*}
	which finishes the proof.
\end{proof}

\vspace{1em}

\begin{proof}{of Theorem {\color{red}2}.}
	First, since the composite log-likelihood $\ell_C$ is smooth in $\gamma$ in a finite Euclidean space and the parameter space $\Gamma$ inherited the norm from the same Euclidean space, so $\ell_{C}$ is continuous in $\Gamma$ and is differentiable in the interior of $\Gamma$. Besides, since $\Gamma$ is compact, so for an arbitrary sample point, and an arbitrary parameter point $\dot{\gamma}$ with $\ell_{C}(\dot{\gamma})>-\infty$, the set
	\begin{equation*}
	\Gamma(\dot{\gamma}) \triangleq \big\{   \gamma\in\Gamma: \ell_C(\gamma) \geq \ell_{C}(\dot{\gamma})     \big\}
	\end{equation*}
	is bounded. For any convergence sequence $\{ \ddot{\gamma}_n \in \Gamma(\dot{\gamma}) \}_{n\geq 1}$, with $\ddot{\gamma}_n \rightarrow \ddot{\gamma}$, we have $\ddot{\gamma} \in \Gamma$ since $\Gamma$ is compact and each $\ddot{\gamma}_n \in \Gamma(\dot{\gamma}) \subset \Gamma$; also, $ \ell_C(\ddot{\gamma}_n) \geq \ell_{C}(\dot{\gamma}) $ for each $n$ implies $\ell_C(\ddot{\gamma}) \geq \ell_{C}(\dot{\gamma}) $. Therefore, $\ddot{\gamma} \in \Gamma(\dot{\gamma})$ and according to Heine-Borel theorem, we conclude that $\Gamma(\dot{\gamma})$ is also a compact set for an arbitrary sample point and parameter point $\dot{\gamma}$.	
	Second, with an arbitrary but fixed sample point, define $\Gamma_{s}$ as the set of stationary point generated by function $M(\cdot) \triangleq \arg\max_{\gamma \in \Gamma} Q (  \gamma \vert \cdot )$, i.e.,
	\begin{equation*}
	\Gamma_{s} = \Big\{  \gamma_s \in \Gamma:  \gamma_s =M( \gamma_s )   \Big\} =  \Big\{  \gamma_s \in \Gamma:  Q ( \gamma_s \vert \gamma_s) = \max_{\gamma \in \Gamma} Q(  \gamma \vert \gamma_s )   \Big\} 
	\end{equation*}
	Since function $Q(x \vert y)$ is smooth in both $x$ and $y$, so for an arbitrary convergence sequence $\{ \gamma_{s,n} \in \Gamma_s \}_{n\geq 1}$ with $\gamma_{s,n} \rightarrow \gamma_{s,0}$, and arbitrary $\gamma$, we have
	\begin{equation}
	Q (\gamma \vert \gamma_{s,0}) = \lim Q (\gamma \vert \gamma_{s,n}) \leq  \lim Q (\gamma_{s,n} \vert \gamma_{s,n}) = Q(\gamma_{s,0} \vert \gamma_{s,0}).  \label{Q1} \tag{15}
	\end{equation}
	Because $\gamma$ is arbitrary in (\ref{Q1}), so $\gamma_{s,0}\in \Gamma_{s}$ and hence $\Gamma_{s}$ is compact. Further we conclude that, $M(\cdot)$ is closed over the complement of $\Gamma_{s}$ since $M(\cdot)$ is identical mapping over the compact set $\Gamma_{s}$. 	
	Third, with the arbitrary but fixed sample point, $\ell_{C}$ is a continuous map on $\Gamma$, and for an arbitrary point $\gamma \in \Gamma_s$, we have $\ell_{C}( M(\gamma) ) = \ell_{C}(\gamma )$, while for an arbitrary point $\gamma \notin \Gamma_s$, we have $Q( M(\gamma) \vert \gamma )> Q( \gamma \vert \gamma )$, hence
	\begin{equation*}
	\ell_{C}(M(\gamma)) - \ell_{C}(\gamma) \geq Q( M(\gamma) \vert \gamma ) - Q( \gamma \vert \gamma )>0.
	\end{equation*}
	Combine the above three conclusions, the global convergence theorem in \cite{wu1983convergence} and \cite{willard1969zangwill} implies that for an arbitrary sample point, our EM-algorithm estimates $\gamma^{(k)}$ converges to a stationary point $\gamma_s$ in the stationary set $\Gamma_{s}$ induced by the function $M(\cdot) \triangleq \arg\max_{\gamma \in \Gamma} Q (  \gamma \vert \cdot )$ when $k\rightarrow \infty$, and $\ell_{C}(\gamma^{(k)})$ converge monotonically to $\ell_{C}(\gamma_s)$.	
\end{proof}

\vspace{1em}

\begin{proof}{of Theorem {\color{red}3}.}
	To avoid confusion, we denote $E_{\gamma}(\cdot)$ as the expectation evaluated under the $P_{\gamma}(\cdot)$, the probability measure induced by $\{H_t,I_t\}$ who follows the distribution in the proposed model with parameter $\gamma=(\beta,\theta)$. To distinguish the oracle parameter value, we denote it as $\gamma_0$.	
	\vskip 0.5em
	Now, for fixed infectiousness function $\{\omega_s,s=1,...,\eta\}$ and hospitalization propensity $\{\tilde\omega_s,s=-1,...,\tilde\eta\}$, we have
	\begin{align*}
	\ell_{C} &=\sum_{0\leq r\leq t} \log \mathbb{P} \big(H_r, I_r |\mathcal{F}_{r-1} \big)\\
	&=\sum_{0\leq r\leq t} \log  \sum_{ h_{r-\tilde{\eta},\tilde{ \eta}}, \cdots, h_{r-1,1} } \mathbb{P}\big(H_r, I_r, h_{r-\tilde{\eta},\tilde{ \eta}}, \cdots, h_{r-1,1} |\mathcal{F}_{r-1}\big) \\	
	&=\sum_{0\leq r\leq t} \log  \sum_{ h_{r-\tilde{\eta},\tilde{ \eta}}, \cdots, h_{r-1,1} }\mathbb{P}\Big( Poisson(\tilde{ \omega }_0 R_{r} \Lambda_r ) = H_r-\sum_{s=1}^{\tilde{ \eta}} h_{r-s,s} \Big)   \\
	&  \qquad \;\;    \cdot  \prod_{s=1}^{\tilde{ \eta}} \mathbb{P} \Big( Binomial( I_{r-s},\tilde{ \omega }_s ) = h_{r-s,s} \Big)   \mathbbm{1} \Big(  H_r-I_r	\leq  \sum_{s=1}^{\tilde{ \eta}} h_{r-s,s} \leq H_r  \Big)    \\
	&  \qquad   \;\; 	 \cdot  \mathbb{P}\Big( Poisson((1-\tilde{ \omega }_0) R_{r} \Lambda_r ) = I_r-H_r+\sum_{s=1}^{\tilde{ \eta}} h_{r-s,s} \Big),  \\
	&=\sum_{0\leq r\leq t}\log\Big[ (R_{r})^{I_r}\exp\big( -R_{r} \Lambda_r \big) C_r\Big] \\
	&=\sum_{0\leq r\leq t} \Big[ I_r\log(R_r)-R_r\Lambda_r+\log C_r \Big]
	\end{align*}
	where 
	\begin{equation*}
	C_r= \Lambda_r^{I_r}(1-\tilde{ \omega }_0)^{I_r}\sum_{ h_{r-\tilde{\eta},\tilde{ \eta}}, \cdots, h_{r-1,1} }  \frac{\left(\frac{\tilde{ \omega }_0}{1-\tilde{ \omega }_0}\right)^{H_r-\sum_{s=1}^{\tilde{ \eta}} h_{r-s,s}} \prod_{s=1}^{\tilde{ \eta}} \mathbb{P} \Big(  h_{r-s,s}  | \mathcal{F}_{r-1}\Big)  }{(H_r-\sum_{s=1}^{\tilde{ \eta}} h_{r-s,s})!(I_r-H_r+\sum_{s=1}^{\tilde{ \eta}} h_{r-s,s})!}
	\end{equation*}
	are constants as functions of $\gamma$.
	\vskip 0.5em
	Thus,	
	\begin{equation*}
	\hat{\gamma} = \arg\max_{\gamma \in \Gamma} \sum_{0\leq r\leq t} \Big[ I_r\log(R_r)-R_r\Lambda_r \Big].
	\end{equation*}	
	Meanwhile, under Condition {\color{red}1}, we have
	\begin{equation}
	\frac{t_0}{t}\sum_{s=1}^{t} \{I_s\log(R_s)-R_s\Lambda_s\}\longrightarrow_{a.s.}E_{\gamma_0} \sum_{0\leq r\leq t_0} \Big[ I_r\log(R_r)-R_r\Lambda_r \Big] \label{ExpConEll} \tag{16}
	\end{equation}
	for some $t_0>0$.
	We now demonstrate that the maximum of the right-hand-side of (\ref{ExpConEll}) over $\gamma\in\Gamma$ is attained uniquely at $\gamma=\gamma_0$. Let $\ell_{Q,r}(\gamma)=(R_{r})^{I_r}\exp[-R_{r} \Lambda_r]$. By Jensen's Inequality, for arbitrary $\gamma_1 \in \Gamma$, we have
	\begin{align*}
	\sum_{0\leq r\leq t_0} E_{\gamma_0}\left(\log\frac{\ell_{Q,r}(\gamma_1)}{\ell_{Q,r}(\gamma_0)}\right)&\le \sum_{0\leq r\leq t_0} \log E_{\gamma_0}\left(\frac{\ell_{Q,r}(\gamma_1)}{\ell_{Q,r}(\gamma_0)}\right)=0,
	\end{align*}
	Hence
	\begin{equation*}
	\sum_{0\leq r\leq t_0} E_{\gamma_0}\Big[\log \ell_{Q,r}(\gamma_1) \Big] \leq  \sum_{0\leq r\leq t_0} E_{\gamma_0}\Big[\log \ell_{Q,r}(\gamma_0) \Big] 
	\end{equation*}
	for arbitrary $\gamma_1\in\Gamma$, and the equality holds if and only if $\ell_{Q,r}(\gamma_1)=\ell_{Q,r}(\gamma_0)$ almost surely. We denote $B^c$ to be the zero-measure set of the parameter space $\Gamma$ such that Condition {\color{red}1} holds for all $\omega\in B$. If, on the contrary that $\hat\gamma \nrightarrow \gamma_0$, then there exist a positive constant $\delta>0$ and a subsequence $\{ t_{k1},t_{k2},\cdots, t_{kn} ,\cdots \}$ such that $\mid \hat{\gamma}(t_{ki}) - \gamma_0 \mid \geq \delta$ for $i=1,2,\cdots,n,\cdots$. By the compactness of $\Gamma$, there further exists a sub-subsequence $\{t_{u_{m1}}, t_{u_{m2}},\cdots, t_{u_{mn}},\cdots\}$ such that $\{  \hat{\gamma}_{t_{u_{mi}}} \}_{i\geq 1}$ is a convergence sub-subsequence and $\hat\gamma (t_{u_{mn}}) \rightarrow \gamma^{\ast}$ for some $\gamma^{\ast}\neq\gamma_0$. Then,
	
	\begin{align*}
	\lim_{n\rightarrow\infty}\frac{t_0}{t_{u_{mn}}}\sum_{s=1}^{t_{u_{mn}}} \log \ell_{Q,r}(\hat\gamma(t_{u_{mn}})) &\ge \lim_{n\rightarrow\infty}\frac{t_0}{t_{u_{mn}}}\sum_{s=1}^{t_{u_{mn}}} \log \ell_{Q,r}(\gamma_0)=E_{\gamma_0}\sum_{s=1}^{t_0}\log \ell_{Q,r}(\gamma_0)\\
	&>E_{\gamma_0}\sum_{s=1}^{t_0}\log \ell_{Q,r}(\gamma^\ast)=\lim_{n\rightarrow\infty}\frac{t_0}{t_{u_{mn}}}\sum_{s=1}^{t_{u_{mn}}} \log \ell_{Q,r}(\gamma^\ast) \\
	&=\lim_{n\rightarrow\infty}\frac{t_0}{t_{u_{mn}}}\sum_{s=1}^{t_{u_{mn}}} \log \ell_{Q,r}(\hat\gamma(t_{u_{mn}})),
	\end{align*}
	which is a contradiction. Hence, $\hat\gamma\rightarrow \gamma_0$ for each $\omega\in B$, i.e., the strong consistency holds for $\hat\gamma$. 	
\end{proof}

\vspace{1em}

\begin{proof}{of Theorem {\color{red}4}.}
	Define 
	\begin{equation*}
	U_t(\gamma)= \sum_{0\leq r\leq t} \frac{\partial \log \ell_{Q,r}(\gamma)}{\partial \gamma} = \sum_{0\leq r\leq t}  \left(\frac{\partial R_r}{\partial\gamma}\right)^TR_r^{-1}(I_r-R_r \Lambda_r)\triangleq \sum_{0\leq r\leq t}  \xi_r(\gamma),
	\end{equation*}
	and let
	\begin{equation*}
	A_t^2= \sum_{0\leq r\leq t}  \text{Cov}(\xi_r(\gamma_0) \mid \mathcal{F}_{r-1} )
	\end{equation*}
	Then the proofs are identical to those of theorem 4.1 in \cite{shi2022robust}, so we only provide an outline of the proof here.
	\vskip 0.5em 
	Since $\hat\gamma$ satisfy $U_t(\hat{\gamma})=0$, so by Taylor expansion, we have
	%For some $\tilde\gamma$ lays between $\gamma_0$ and $\hat\gamma$, we have	
	\begin{align}
	&0=U_t(\hat\gamma)=U_t(\gamma_0)+\frac{\partial U_t(\gamma)}{\partial\gamma}|_{\gamma=\tilde\gamma}(\hat\gamma-\gamma_0), \quad i.e.,  \\
	&-A_t^{-1}\frac{\partial U_t(\gamma)}{\partial\gamma}|_{\gamma=\tilde\gamma}(A_t^{-1})^T A_t(\hat\gamma-\gamma_0)=A_t^{-1}U_t(\gamma_0). \label{Ta} \tag{17}
	\end{align}	
	For an arbitrary given and fixed vector $\alpha=(\alpha_1,...,\alpha_{p+2})^T\in \mathcal{M}_{(p+2)\times 1}$, define
	\begin{align*}
	S_t(\alpha) \triangleq  \frac{ \alpha^T}{\Vert \alpha \Vert} A_t^{-1} U_t(\gamma_0) &= \sum_{0\leq r\leq t} \frac{ \alpha^T}{\Vert \alpha \Vert} A_t^{-1} \left(\left(\frac{\partial R_r}{\partial\gamma}\right)^T R_r^{-1}(I_r- R_r \Lambda_r) \right)  \triangleq \sum_{0\leq r\leq t}  \tilde{\xi}_r .
	\end{align*}
	It's trivial to see that $\tilde{\xi}_r \in \mathcal{F}_r$ and $\mathbb{E} (\tilde{\xi}_r | \mathcal{F}_{r-1})=0$, hence $\{  \tilde{\xi}_r, 0\leq r\leq t \}$ is a martingale difference sequence. So
	\begin{align*}
	\tilde{V}_t^2 (\alpha) &\triangleq \sum_{0\leq r\leq t}   \mathbb{E} \big(  \tilde{\xi}_r^2 | \mathcal{F}_{r-1}    \big)=\frac{ \alpha^T}{\Vert \alpha \Vert} A_t^{-1} V_t^2(\gamma_0) ( A_t^{-1} )^T  \frac{ \alpha}{\Vert \alpha \Vert}
	\end{align*}
	with $V_t^2(\gamma) \triangleq \sum_{0\leq r\leq t}   \mathbb{E} \big(  \xi_r^2(\gamma) | \mathcal{F}_{r-1}    \big)$, is the conditional variance that would affect the limiting distribution of $ S_t(\alpha)$. Under Condition {\color{red}2}, we have 
	\begin{equation*}
	\tilde{V}_t^2 (\alpha)\longrightarrow_{P}\zeta^2(\alpha)
	\end{equation*}
	Together with Condition {\color{red}3}, we conclude that $\tilde{V}_t^2(\alpha) \longrightarrow_{L_1} \zeta^2(\alpha)$ and $\mathbb{E}  \tilde{V}_t^2 (\alpha) = \mathbb{E}  \zeta^2(\alpha)  =1 $. Hence $\mathbb{E}(\max_{0\leq r\leq t}\tilde{\xi}_r^2 ) \leq 1$. Combining the conditional Lindeberg condition (Condition {\color{red}4}), we conclude $\tilde{U}_t^2 (\alpha) - \tilde{V}_t^2(\alpha) \overset{L_1}{\longrightarrow} 0$ by theorem 2.23 of \cite{hall2014martingale} for $\tilde{U}_N^2 (\alpha)\triangleq \sum_{0\leq r\leq t } \tilde{\xi}_r^2$, hence $\tilde{U}_t^2(\alpha) \overset{L_1}{\longrightarrow} \zeta^2(\alpha)$.
	\vskip 0.5em
	Thus, by using the martingale central limit theorem (see e.g, theorem 3.2 of \cite{hall2014martingale}), $A_t^{-1}U_t(\gamma_0)$ converge to certain distribution with characteristic function
	\begin{equation*}
	f_{A_t^{-1}U_t(\gamma)} (\alpha)  =\mathbb{E} \exp\Big(  i\alpha ^T A_t^{-1} U_t(\gamma_0)   \Big) \rightarrow \mathbb{E}\exp \Big(  -\frac{1}{2} \Vert \alpha \Vert_2^2 \zeta^2(\alpha)      \Big),
	\end{equation*}
	for arbitrary $\alpha$,or equivalently,
	\begin{equation*}
	A_t^{-1}U_t(\gamma_0)  \longrightarrow_d \zeta^T \cdot Z  \;\; ({\rm stably}),  \label{distribution}
	\end{equation*}
	for $\zeta$ independent of $Z\sim \mathcal{N}(0,I)$. Of course, the limit distribution holds only on the non-extinction set $\mathcal{E}_{none}$. Since $\zeta$, $Z$ are also measurable, so we have
	\begin{equation}
	A_t^{-1}U_t(\gamma_0) \longrightarrow_P \zeta^T \cdot Z. \label{normal} \tag{18}
	\end{equation}
	\vskip 0.5em
	Now, combining Condition {\color{red}5} and Theorem {\color{red}3} insures $\hat\gamma \rightarrow_{a.s.} \gamma_0$, i.e., $\tilde{\gamma}$ would be in the neighborhood of $\gamma_0$, so we have 
	\begin{equation}
	A_t^{-1}  \left( - \frac{\partial U_t(\gamma)}{\partial \gamma}\Big|_{\gamma= \tilde{\gamma} }   \right) (A_t^{-1})^T  \overset{P}{\longrightarrow}  \zeta^T\zeta .  \label{weight} \tag{19}
	\end{equation}
	Thus, by (\ref{Ta}), (\ref{normal}), (\ref{weight}), and our choice of $A_t^2$, we conclude that,
	\begin{equation*}
	\Big[ \sum_{1\leq r\leq t}\text{Cov}  \big(   \xi_r(\gamma_0) | \mathcal{F}_{t-1}    \big)	  \Big]^{1/2} \big(  \hat{\gamma} -\gamma_0    \big) \longrightarrow_d N(0,I),
	\end{equation*}
	which finishes the proof.
\end{proof}

\subsection{Additional results from the simulation study and application to COVID-19 Data}
\renewcommand{\thefigure}{S7}
\begin{figure}[!p]
	\begin{center}
		\includegraphics[width=1\textwidth]{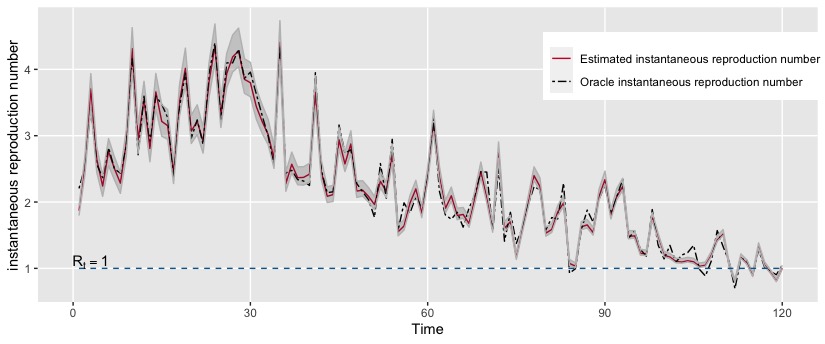}
		\caption{Estimation of the daily instantaneous reproduction number, $R_t$, when the reported number of daily new infections was accurate, the infectiousness function $\omega_s$ and hospitalization propensity $\tilde\omega_s$ were unknown and no prior knowledge was available in the simulation study. Black dotted line stands for the oracle $R_t$. Red solid line and the grey shadow stand for the estimates and its corresponding Bootstrapping confidence interval ($90\%$ confidence level) using the proposed method.}
		\label{fig:EstR}
	\end{center}
\end{figure}

\renewcommand{\thefigure}{S8}
\begin{figure}[!p]
	\begin{center}
		\includegraphics[width=1\textwidth]{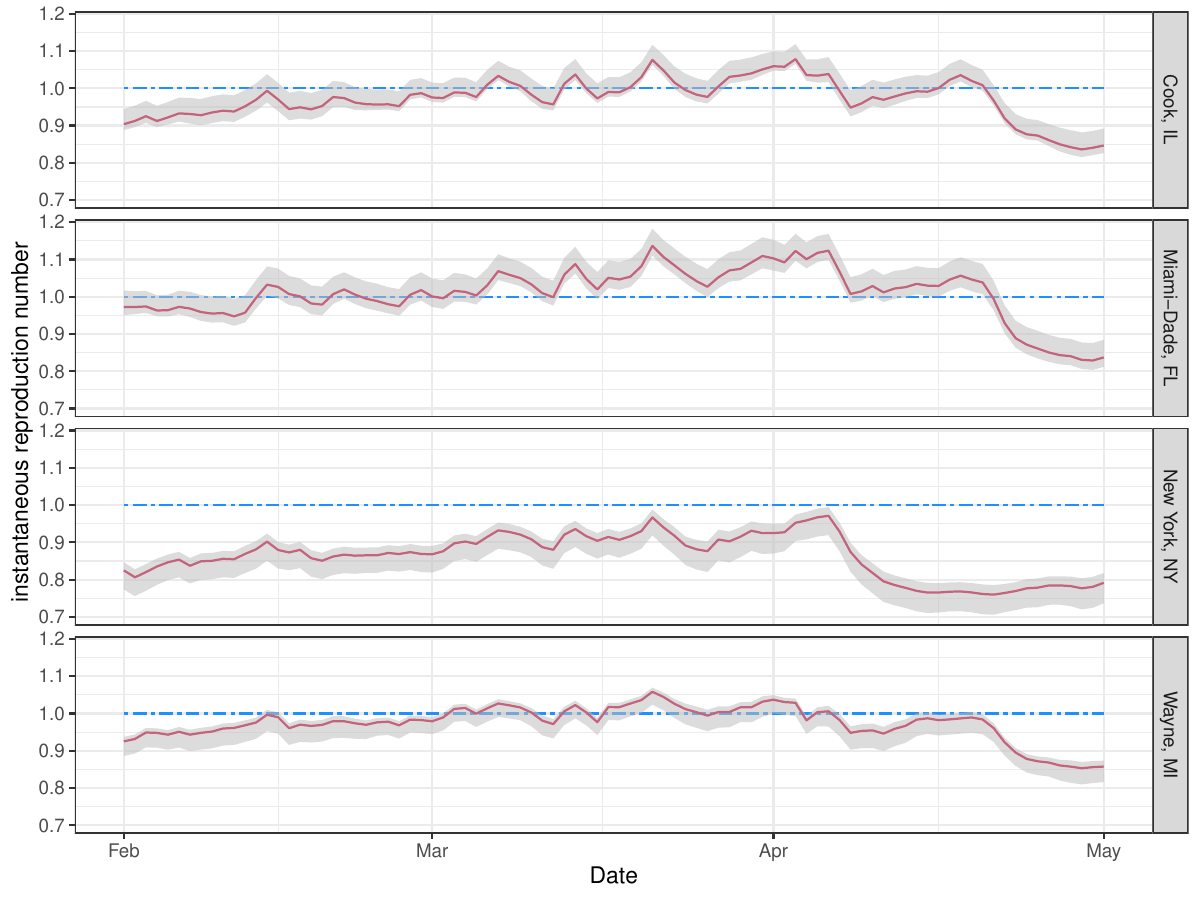}
		\caption{Estimated instantaneous reproduction number, $R_t$, for COVID-19 in the four studied counties during February to May 2021. The blue lines represent the critical value of $R_t=1$. Red lines represent the estimated county-level $R_t$ during the study period. Shaded areas indicate the bootstrap confidence intervals ($90\%$ confidence level).
		}
		\label{county level: estimates}
	\end{center}
\end{figure}

\renewcommand{\thefigure}{S9}
\begin{figure}[!p]
	\begin{center}
		\includegraphics[width=1\textwidth]{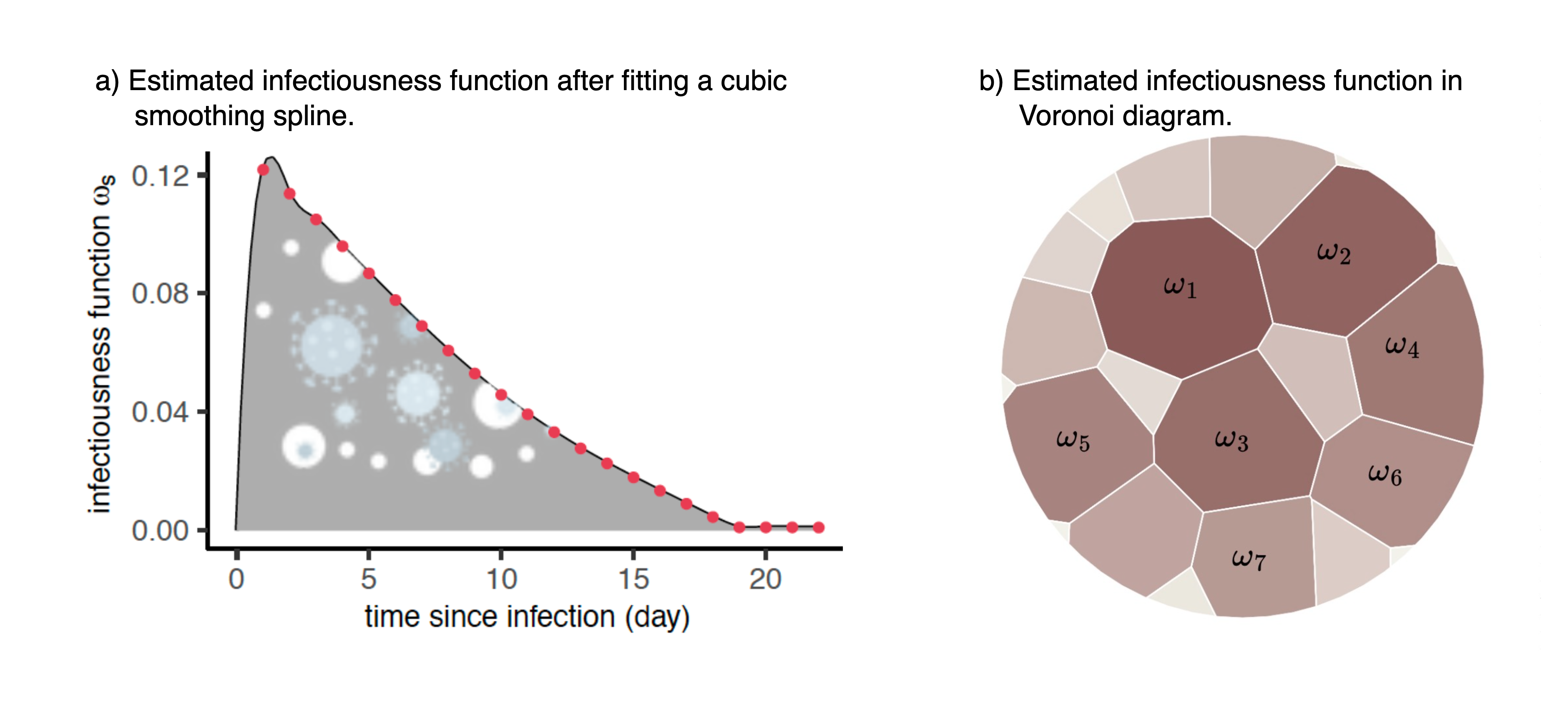}
		\caption{Visualization of the estimated infectiousness function, $\omega_s$, for COVID-19 in the four studied counties during February to May 2021. In panel a), the red dots represent the estimated $\omega_s$, $1\leq s\leq 22$, while the black line represents the estimated infectiousness function after fitting a cubic smoothing spline. In panel b), the estimated $\omega_s$ values are visualized using a Voronoi diagram to compare the magnitude of each $\omega_s$. Values of $\omega_s$ in the first week since infection are labeled.}
		\label{fig:Case5Omega}
	\end{center}
\end{figure}

\bibliographystyle{agsm.bst}

\bibliography{ref}